\documentclass[aps,pra,reprint]{revtex4-1}

\usepackage{graphicx}
\graphicspath{{images/}{images/uni-quantum}{images/uni-classical}{images/bi-classical}{Images2PDF/}} 
\usepackage{subfigure}
\usepackage{float}
\usepackage{picture}

\usepackage{dcolumn}
\usepackage{bm}
\usepackage[usenames,dvipsnames]{xcolor}
\definecolor{dark-gray}{gray}{0.3}
\usepackage{hyperref}
\hypersetup{
	pdftitle={},
	pdfauthor={Alexander Roth, Klemens Hammerer},
	pdfcreator={Alexander Roth, Klemens Hammerer},
	pdfproducer={Alexander Roth, Klemens Hammerer},
	pdfsubject={},
	pdfkeywords={},
	colorlinks,
	bookmarksopen = true,
	bookmarksnumbered = true,
	citecolor=Blue,
	linkcolor=Blue,
	urlcolor=Blue
 	}

\usepackage{amsmath, amscd, amsfonts, amssymb, amsthm} 
\usepackage{ textcomp }
\usepackage{mathtools}
\usepackage{dsfont}
\usepackage[arrow, matrix, curve]{xy}
\usepackage{ifthen}

\newcommand{\B}[1]{\left(#1\right)}		
\newcommand{\C}[3][]{\left[ #2,\, #3\right]\onlyifnotempty{#1}{_{#1}}}   
\newcommand{\Curly}[1]{\left\{#1\right\}}		
\newcommand{\ket}[1]{\left|#1 \right\rangle}		
\newcommand{\Mean}[1]{\left\langle#1\right\rangle}		
\newcommand{\Meansmall}[1]{\langle#1\rangle}		

\newcommand{\Exp}[1]{\exp\B{#1}}

\newcommand{\sumsetclap}[1][]{\sum_{\mathclap{\substack{#1}}}}

\newcommand{\fracGeneral}[4][]{\onlyifnotempty{#1}{\left.}\frac{#2 #3}{#2 #4} \onlyifnotempty{#1}{\right|_{#1}} }

\newcommand{\D}[2][]{\fracGeneral[#1]{d}{}{#2}  }

\newcommand{\smallmatrixx}[1]{\left(\begin{smallmatrix}#1\end{smallmatrix}\right)}
\newcommand{\matrixx}[1]{\begin{pmatrix}#1\end{pmatrix}}

\newcommand{\Kappa}{{\tilde\kappa}}

\newcommand{\define}[2][]
{\ifthenelse{\equal{#1}{}}
{\textbf{\textit{#2}}\label{def:#2}\index{#2}\xspace}
{\textbf{\textit{#1}}\label{def:#2}\index{#2@#1}\xspace}}	

\newcommand{\oo}{\infty}
\newcommand{\intd}[1]{\mathrm{d}#1 \; }

\newcommand{\Min}[1]{\text{Min}\B{#1}}

\newcommand{\dg}{^\dagger}
\newcommand{\cc}{^*}

\renewcommand{\1}{\mathds{1}}

\newcommand{\breaksign}[1]{ \notag\\  & \phantom{=}\, #1 }

\renewcommand{\L}[2][\rho]{{\mathcal D}\left[#2\right] #1}

\renewcommand{\a}{{\hat a}}
\renewcommand{\b}{{\hat b}}
\renewcommand{\c}{{\hat c}}

\newcommand{\s}{\hat\sigma}

\newcommand{\ad}{\hat a^\dagger}
\newcommand{\bd}{\hat b^\dagger}

\newcommand{\onlyifnotempty}[2]{\ifthenelse{\equal{#1}{}}{}{#2}}

\renewcommand{\Re}[1]{\,\text{Re}\left[#1\right]}
\newcommand{\Resmall}[1]{\,\text{Re}[#1]}

\newcommand{\bracketsize}[1]{
\ifthenelse{\equal{#1}{1}}
    {\big}
    {\ifthenelse{\equal{#1}{2}}
	{\Big}
	{\ifthenelse{\equal{#1}{3}}
	    {\bigg}
	    {\ifthenelse{\equal{#1}{4}}
		{\Bigg}
		{}	    
	    }
	}
    }
}


\begin{document}

\title{Synchronization of Active Atomic Clocks via Quantum and Classical Channels}

\author{Alexander Roth}
\email{alexander.roth@itp.uni-hannover.de}
\affiliation{Institute for Theoretical Physics, Institute for Gravitational Physics (Albert Einstein Institute), Leibniz University Hannover, Callinstra{\ss}e 38, 30167 Hannover, Germany}
\author{Klemens Hammerer}%
\affiliation{Institute for Theoretical Physics, Institute for Gravitational Physics (Albert Einstein Institute), Leibniz University Hannover, Callinstra{\ss}e 38, 30167 Hannover, Germany}


%

\date{\today}

\begin{abstract}
\noindent   Superradiant lasers  based on  atomic ensembles exhibiting ultra-narrow optical transitions can emit light of unprecedented spectral purity and may serve as active atomic clocks.  We consider two frequency-detuned active atomic clocks, which are coupled in a cascaded setup, i.e. as master \& slave lasers, and study the synchronization of the slave to the master clock. In a setup where both atomic ensembles are coupled to a common cavity mode such synchronization phenomena have been predicted by Xu et al. [Phys. Rev. Lett. {\bf 113}, 154101 (2014)]  and experimentally observed by Weiner et al. [arXiv:1503.06464 (2015)].  Here we demonstrate that synchronization still occurs in cascaded setups but exhibits distinctly different phase diagrams. We study the characteristics of synchronization in comparison to the case of coupling through a common cavity. We also consider synchronization through a classical channel where light of the master laser is measured phase sensitively and the slave laser is injection locked by feedback and compare to the results achievable by coupling through quantum channels.
%
%
\end{abstract}

\pacs{Valid PACS appear here}
\maketitle


\section{Introduction}
\label{sec:introduction}

Atomic clocks based on optical transitions already achieve record precisions with fractional uncertainties of $10^{-18}$ \cite{Nicholson2015} and offer great potential for further improvements \cite{Ludlow2015}. Notably, current optical clocks are limited in precision by the instability of the laser used for interrogating the atomic reference system rather than by the linewidth of the clock transition \cite{Al-Masoudi2015}. In order to overcome this limitation the concept of an active atomic clock has been suggested where a  lattice of cold atoms with ultra-narrow clock transition itself serves as a laser gain medium resulting in radiation with extremely narrow linewidth in the mHz regime \cite{meiser_prospects_2009,meiser_intensity_2010,bohnet_steady-state_2012,bohnet_active_2013,Norcia2016}. This would remedy the need to reference an external laser to an atomic clock transition.

An active clock laser operates in regime with inverted timescales as compared  to a normal laser \cite{meiser_prospects_2009}:
In the usual case  atoms are pumped incoherently faster than the laser cavity decays.  The cavity amplitude then amplifies through  stimulated emission  only those frequencies which fit within the cavity linewidth.   In an active clock laser the atoms are are  pumped incoherently much slower than the cavity decays. Due to the long lifetime of atomic coherences correlations between the atoms build up  giving to a collectively enhanced, superradiant emission into the cavity. The correlations between the atoms result in  a linewidth  of the output light which is on the order of the one of the atomic transition itself. 

Such a superradiant laser exhibits further remarkable properties: Recently it was shown by Xu et al.~\cite{xu_synchronization_2014} and experimentally demonstrated by Weiner et al.~\cite{weiner_phase_2015} that two frequency-detuned atom-ensembles coupling to the same cavity mode operated in the superradiant regime synchronize in a large parameter regime; they radiate at the mean frequency while preserving the narrow linewidth.  For larger detuning the ensembles will cross through a phase transition separating the synchronized from the unsynchronized phase and then behave like  two independent superradiant lasers at their natural frequency. The synchronization dynamics of superradiant lasers serving as active atomic clocks receives particular importance in the perspective of quantum networks of atomic clocks as envisioned in \cite{Komar2014} for enhanced positioning, navigation and geodesy. However, the results of  \cite{xu_synchronization_2014,weiner_phase_2015} cannot be directly applied to the context of synchronization of remote atomic clocks  as the two atomic ensembles are coupled to a common cavity mode.

In the present work we extend the analysis of \cite{xu_synchronization_2014} and consider two remote superradiant lasers coupled through an optical channel in the cascaded configuration of a master and a slave laser. We determine the phase diagram of synchronization in a cascaded setup and determine the parallels and differences to the case of a setup with symmetric coupling studied in \cite{xu_synchronization_2014,weiner_phase_2015}. In short, our findings are: synchronization still occurs in a cascaded configuration but the common frequency will always be the one of the master laser in this case. Furthermore, for symmetric coupling the two ensembles  in the synchronized phase radiate as one ensemble with $2N$ atoms. This does not occur so in the cascaded setup resulting in   changes in the synchronization phase diagram. In either case the linewidth of radiation is the same.

The synchronization of quantum systems has been the subject of several theoretical studies lately  \cite{zhirov_quantum_2009,lee_quantum_2013,mari_measures_2013,walter_quantum_2014,lee_entanglement_2014}. Despite these efforts there is no clear measure to distinguish genuine {\it quantum} synchronization of quantum systems and from a {\it classical} synchronization of quantum systems. In order to compare the two scenarios in the present context we further extend our analysis and consider toy models where the two lasers are locked to each other through a classical channel, that is, a measurement and feedback procedure. We study both cascaded and symmetric couplings through (idealized) classical channels and compare to the results achieved through a quantum channel. In particular we discuss the impact on the linewidth.

The article is organized as indicated in the following table:
\begin{table}[H] \centering\renewcommand{\arraystretch}{1.4}
\begin{tabular}{r|c|c}
		  & Symmetric coupling	& Cascaded setup	\\\hline
Quantum channel	&  Review of \cite{xu_synchronization_2014}, Sec.~\ref{sec:quantum bi}	& Sec.~\ref{sec:quantum uni}	\\\hline
Classical channel& Sec.~\ref{sec:classical bi}	& Sec.~\ref{sec:classical uni}	\\
\end{tabular}
\end{table}

\section{Synchronization through Quantum Channels}

\subsection{Synchronization of Two Atomic Ensembles in a Common Cavity}
\label{sec:quantum bi}
\begin{figure}[t]\centering
{\includegraphics[width=.85\linewidth]{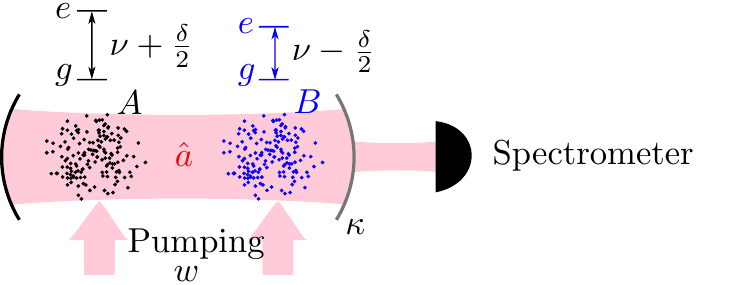}}
 \caption{Two ensembles of two level systems ($A $ and $B $) coupling to the same cavity mode $\a $, as considered in \cite{xu_synchronization_2014}.  The frequencies of the transitions $\ket g\leftrightarrow\ket e$ are detuned by $\pm\delta/2$ from the cavity resonance at frequency $\nu$ for ensemble $A$ and $B$ respectively. Atoms are pumped incoherently from $\ket g$ to $\ket e$ via a third fast decaying level (not shown in level scheme) at rate $w $ and decay from $\ket e$ to $\ket g$ predominantly through the cavity. The cavity decays at rate $\kappa$.}
 \label{fig:xucavity}
\end{figure}
In this section we will briefly review the  setup, methods,  and results of Xu et al. \cite{xu_synchronization_2014}. We aim to present a sufficient level of detail in order to provide a self-contained derivation of the results going beyond the work of Xu et al. in the later sections. For more details we refer to the excellent presentation in \cite{xu_synchronization_2014}. The setup  in Fig.~\ref{fig:xucavity} consists of two ensembles of atoms $A$ and $B$, each containing $N$ atoms, placed in the same cavity. Atoms are assumed to have two relevant internal levels $\ket g$ and $\ket e $. The transition frequencies of atoms in ensemble $A$ and $B$ have a relative frequency detuning of $\delta $ while all atoms within each ensemble are assumed to be frequency degenerate. The transition $\ket g\leftrightarrow\ket e $ couples to the cavity mode $\a $ with a single photon Rabi frequency $\Omega/2 $.  Ensemble $A$ is detuned from the cavity resonance by $\delta/2$ and ensemble $B$ by $-\delta/2$. The cavity linewidth is $\kappa$, and we will ultimately assume the bad cavity limit such that the assumptions regarding the detuning of atoms from cavity resonance are insignificant. Atoms decay from $\ket e$ to $\ket g $ into free space  with rate $\gamma_s $ and dephase with rate $T_2^{-1} $, and at the same time they are incoherently repumped   from $\ket g$ to $\ket e $ with the rate $w $   (e.g. through an already eliminated third level). In a rotating frame at the cavity frequency the system is described by the Lindblad master equation
\begin{align}
\dot\rho &=  -i
    \C{\frac{  \delta }{2}   \B{ \hat J_A^z -\hat J_B^z   } +  \frac{  \Omega }{2}  \B{\ad (\hat J_A^-+\hat J_B^-) + \mathrm{h.c.}}
    }{\rho}\nonumber  \\
  &\quad
+\kappa \L{\a}+ \sumsetclap[T =A,B\\j=1\dots N] ~ \B{\gamma_s \L[]{\s^-_{T,j}}   +w \L[]{\s^+_{T,j}} 
   }     \rho	.
\label{eq:bi-quantum master equation}
\end{align}
$\s^z_{T,j}$ and $\s^\pm_{T,j}$ are the usual Pauli matrices for the $\ket g\leftrightarrow\ket e $ transition for atom $j\in\{1\ldots N\}$ in ensemble $T\in\{A,B\}$. We use the collective spin operators   
$ \hat J_T^\pm:= \sum_{i=1}^N \s_{T,i}^\pm$, $ \hat J_T^z:=\frac 1 2  \sum_{i=1}^N \s_{T,i}^z$, and the Lindblad superoperator  \mbox{$\L{A} :=  A \rho {A}\dg  - \frac 1 2 \C[+]{ {A}\dg A} \rho   $}.
Steady state superradiance is achieved with a dominating cavity decay  $\kappa\gg w $, which is inverted  compared to an ordinary laser where the pumping  dominates  $w \gg \kappa  $. The fast decay of the cavity with rate $\kappa $ compared to all other timescales in the system  allows for an adiabatic elimination of the cavity mode
\begin{align}
\a \simeq   -\frac{i\Omega}{\kappa+i \delta } \hat J_A^- -\frac{i\Omega}{\kappa -i \delta} \hat J_B^- \approx-\frac{i\Omega}{\kappa } (\hat J_A^-		+\hat J_B^-)
\label{eq:qu-bi adiabatic elimination}
,
\end{align}
where we used the approximation that the detuning \mbox{$\delta \ll \kappa $} is small  compared to the cavity linewidth. After adiabatic elimination the   decay of the cavity  $\kappa \L{\a}$  translates to a collective decay of the atoms $ {\gamma} \mathcal D  [ \hat J^- ]\rho $ at rate  ${\gamma}= {\Omega^2}/{\kappa} $. The decay into the cavity mode is enhanced by a factor of  $N $  and dominates  the decay process  \cite{meiser_intensity_2010}, i.e. $ {\gamma} N \gg \gamma_s,T_2^{-1}  $, allowing us to drop the emission into free space and the dephasing
\begin{align}
   \dot\rho  &=   - \frac{i\delta}{2  }       \C{ \hat J_A^z -\hat J_B^z  }{\rho}        +      {\gamma}  \L{\hat J_A^- +J_B^-  }\nonumber\\
     &\quad+  \sumsetclap[T =A,B\\j=1\dots N] ~  w \L{\s^+_{T,j}}.
   \label{eq:qu-bi eliminated}
 \end{align}

This dynamics can be solved in a mean field approximation with respect to the mean polarization of atoms along $z$, as developed in Refs.~\cite{xu_synchronization_2014} and \cite{meiser_prospects_2009,meiser_intensity_2010}. Due to the symmetry of \eqref{eq:qu-bi eliminated} all expectation values of Pauli operators must be  symmetric under exchange of the particles in each ensemble. Additionally, the differential equations of the expectation values involving only one ensemble are independent of $\delta $ and therefore identical for both ensembles, allowing us to drop unnecessary indices  $\Meansmall{\s^ {\pm,z}} =\Meansmall{\s_{A,i}^ {\pm,z}} $,   $\Meansmall{\s_1^+\s_2^- }= \Meansmall{\s_{A,i}^+\s_{A,j}^- }= \Meansmall{\s_{B,i}^+\s_{B,j}^- } ~ \forall i\neq j$ and $ \Meansmall{\s_A^+\s_B^- }= \Meansmall{\s_{A,n}^+\s_{B,m}^- }   ~ \forall n,m$. Exploiting the  symmetry of the master equation  it holds  $\Meansmall{{\s_{T,i}^\pm}}=0 $. In order to arrive at a closed set of differential equations third order cumulants are set to zero \cite{meiser_intensity_2010,xu_synchronization_2014} factorizing third order moments into  first and second order moments. Additionally we approximate $ \Meansmall{\s_{A,1}^ {z}\s_{A,2}^ {z}} \approx \Meansmall{\s^ {z}}^2 $, which holds true outside of the regime  of very weak pumping  $w<{\gamma_{}}, T_2^{-1}, \gamma_s $ \cite{xu_synchronization_2014}. The mean polarization in stationary state in leading order $1/N $  is found to be
\begin{align}
\Mean{\s^z} &=\begin{cases}
                 \min\left(\dfrac{w^2+\delta^2}{2 w N\gamma},1\right) ,&   0\leq \delta< w \\[1em]
		        \min\left(\dfrac{w }{ N\gamma} ,1\right),&   \delta\geq w
               \end{cases}
               \label{eq:quantum bi sol z}
              .
\end{align}

\begin{figure}[t]\centering
{\includegraphics[width=.95\linewidth]{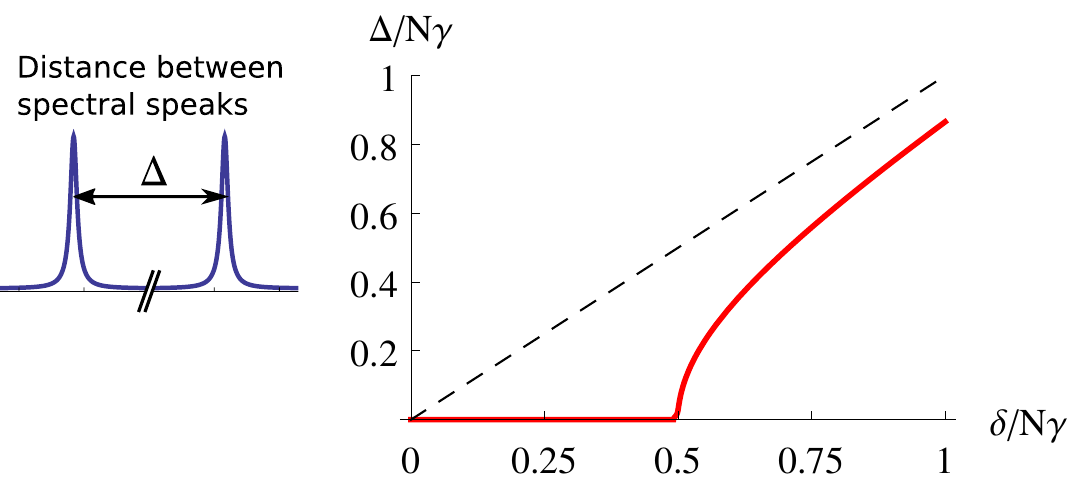}}
 \caption{Effective detuning $\Delta $ between the spectral peaks of light emerging from the laser cavity versus detuning $\delta$ between the bare transition frequencies of the two ensembles. For $\delta<w$, the rate of incoherent pumping of atoms, the two peaks coalesce signifying synchronization of atoms. The dashed line is $\Delta=\delta $ and is approached asymptotically for $\delta\gg w$.}
 \label{fig:xu3}
\end{figure}

The synchronization of the two ensembles is witnessed by the spectrum of light emitted from the cavity which is given by the Fourier transform of the two-time correlation function $\Meansmall{ \ad(\tau) \a(0)}$ of the intra-cavity field. In view of Eq.~\eqref{eq:qu-bi adiabatic elimination} this requires evaluation of the two-time correlations of atomic dipoles, which can be done by means of the quantum regression theorem. For later reference we  explicitly state the corresponding equations of motion for atomic two-time correlation functions,
 \begin{align}
  \D{\tau} \smallmatrixx{
  \Mean{ \s_A^+(\tau) \s_B^-(0)}
  \\
   \Mean{  \s_1^+(\tau)\s_2^-(0) }} &=\frac 1 2  \matrixx{X & Y \\Y& X\cc}   \smallmatrixx{
  \Mean{ \s_A^+(\tau) \s_B^-(0)}
  \\
   \Mean{  \s_1^+(\tau)\s_2^-(0) }}
   \label{eq:bi-quantum dynamics of expectation values}
 \end{align}
 where  $ X=     {\gamma}   ( N-1) \Meansmall{ \s^z} - \gamma -w +i \delta$, and $  Y=    N\gamma \Meansmall{\s^z}$, cf. Eq.~(8) in \cite{xu_synchronization_2014}. The two-time correlation functions, the solution of \eqref{eq:bi-quantum dynamics of expectation values},    consists of  linear combinations of $ \Exp{-\B{\Gamma_0 \pm x_0}\tau/2} $, where  $ \Gamma_0:= w - \gamma  (N-1)  \Meansmall{\s^z}+\gamma$, and $  x_0:=   \sqrt{  (N\gamma    \Meansmall{\s^z})^2-\delta ^2} $. For $\delta\gg w$ this corresponds to two components oscillating at frequencies $\pm\delta/2$ and decaying at rate $\Gamma_0$. The spectrum thus consists of two separate peaks of width $\Gamma_0$ at the bare transitions frequency $\nu\pm\delta/2$ of each ensemble. For smaller detuning $\delta$ the coupled dynamics of the two ensembles of atoms first exhibits frequency pulling giving rise  an effective detuning $\Delta<\delta$ between the two peaks as long as $\delta> w $, cf. Fig.~\ref{fig:xu3}. For $\delta<w $ the two peaks merge and the two ensembles radiate at the same frequency signifying synchronization. The corresponding widths are given by
  \begin{align}
 \Gamma/{\gamma} &= \begin{cases}
                 \dfrac{w^2+\delta^2}{2 wN {\gamma} }+1 ,&   0\leq \delta< w \\[1em]
		 \dfrac{w }{ N{\gamma}}+1 ,&   \delta\geq w
               \end{cases}
               \label{eq:quantum bi linewidth}
 \end{align}
in the superradiant regime, which is upper bounded by $\Meansmall{\s^z} <1 $ using \eqref{eq:quantum bi sol z}.

\begin{figure}[t]\centering
{\includegraphics[width=.771\linewidth]{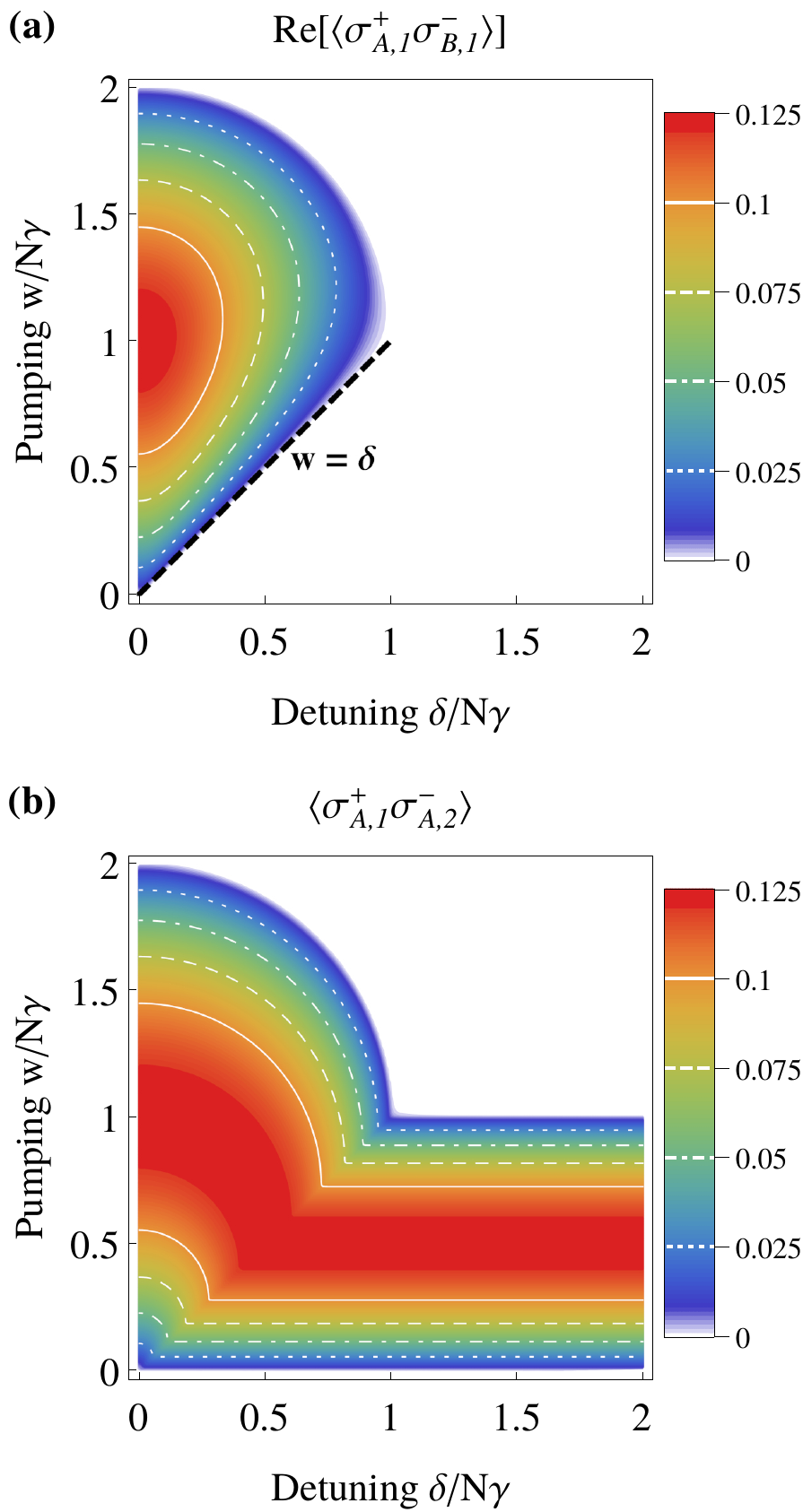}}
  \caption{ (a) Non-vanishing inter-ensemble correlations  $\Resmall{\Meansmall{\s_A^+ \s_B^-}}   $  outline the synchronized parameter    regime. The dashed line $w=\delta $ separates the synchronized from the  unsynchronized superradiant regime.
   {(b)} shows the inner-ensemble correlations  $\Meansmall{ \s_1^+  \s_2^-}   $ equal for both ensembles.    For detuning smaller than the  incoherent pumping rate $\delta <w$ both ensembles  are synchronized and the critical pumping rate is moved from $w=N\gamma $ for $\delta> w\xi $  to $w=2 N\gamma $ for $\delta=0 $. Both plots use  $N\gamma=10^6 \text{Hz} $.
    }
  \label{fig:bi-quantum contourplotReAB}
\end{figure}


Synchronization of the two active atomic clocks physically means that the collective atomic dipoles oscillate in phase. This corresponds to a large non-zero average value of $\Meansmall{\vec \sigma_A^\perp \cdot \vec \sigma_B^\perp}$ where $\sigma_{A(B)}^\perp$ denotes the spin component transverse to the mean polarization along $z$ for ensemble $A(B)$ (each of which is zero on average in steady state, $\Meansmall{\sigma_{A(B)}^\perp}=0$). It is straight forward to check that  $\Meansmall{\vec \sigma_A^\perp \cdot \vec \sigma_B^\perp}   = 4 \Resmall{  \Meansmall{\s_A^+   \s_B^-} } $ such that these    inter-ensemble   correlations can also be directly used as a measure for synchronization \cite{zhu_synchronization_2015}.   It is instructive to directly look at this quantity in its dependence on the pumping $w$ and the bare detuning $\delta$, see {Fig.~\ref{fig:bi-quantum contourplotReAB}a}. The regime of synchronization is clearly visible as the regime of  non-vanishing inter-ensemble correlations. This regime is bounded by $w=\delta$  and the quarter circle $ (w-N\gamma)^2 +\delta^2 =(N\gamma)^2 $, which can be derived from \eqref{eq:quantum bi sol z}. The synchronization can be understood as nothing else but the transition from two independent superradiant ensembles $\delta\gg w $ to one superradiant  ensemble with $2N $ particles for $\delta=0 $.
For $\delta\gg w $ the superradiance is visible in  non-vanishing   inner-ensemble  correlations  $\Meansmall{\vec \sigma_1^\perp \cdot \vec \sigma_2^\perp} = 4   \Meansmall{\s_1^+   \s_2^-}  $ (see {Fig.~\ref{fig:bi-quantum contourplotReAB}}b) and their independence in vanishing inter-ensemble  correlations.
Decreasing  $\delta $ into the synchronized regime    inter-ensemble  correlations build up, approaching  the inner-ensemble correlations, until $\delta=0 $ where there is no difference between both ensembles and $ {  \Meansmall{\s_A^+   \s_B^-} } =\Meansmall{\s_1^+   \s_2^-}  $.
The additional inter-ensemble correlations in the synchronized regime make the  the collective spin $\vec J_A+\vec J_B $ more robust against   noise and move  the critical pumping rate for the  phase transition between  superradiant emission and chaotic light  to $w=2N{\gamma} $ for $\delta=0 $.


The overall photon flux emerging from the cavity is, for large $N$,
\begin{align*}
\Mean{\ad_\textrm{out} \a_\textrm{out}} &\approx   2 \gamma  N^2  \B{    \Mean{ \s_1^+  \s_2^-  }   +   \Re{ \Mean{ \s_A^+  \s_B^-  }}}
,
\end{align*}
which follows from Eq.~\eqref{eq:qu-bi adiabatic elimination} and the input-output relation $\a_\textrm{out}= \a_\textrm{in}+\sqrt{\kappa} \a $ \cite{gardiner_quantum_2004}.
For $\delta=0 $ the photon flux scales  proportional to $(2N)^2 $, as one would expect of one ensemble with $2N $ atoms, and scales with $2 N^2 $ for two independent ensembles each with $N $ atoms.
\subsection{Two Atomic Ensembles in Separate Cascaded Cavities}
\label{sec:quantum uni}

\begin{figure}[t]\centering
{\includegraphics[width=.95\linewidth]{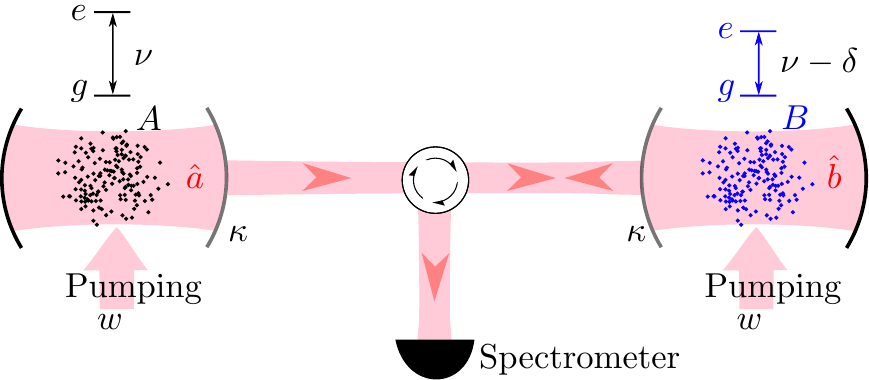}}
  \caption{Two ensembles of two level systems ($A $ and $B $) coupling to the cavity modes $\a,\b $ respectively.  The transition frequencies $\ket g \leftrightarrow \ket e$ of  ensemble $A $ and the cavity frequencies $\a,\b $  are  $\nu $, while   ensemble $B $'s transition frequency  is detuned by $-\delta $.  Atoms are pumped incoherently from $\ket g$ to $\ket e$ via a third fast decaying level (not shown in level scheme) at rate $w $ and decay from $\ket e$ to $\ket g$ predominantly through the cavity.  The  cavities decay with rate $\kappa $  and the output of cavity $\a $ is directly injected into cavity $\b $. The output of cavity $\b $ is diverted into a spectrometer and not into cavity $\a $ using a lossless Faraday rotator.}
  \label{fig:uni-quantum setup}
\end{figure}

Next we are going to consider an alternative setup where the two atomic ensembles are kept in separate cavities which are coupled unidirectionally: Light emerging from the cavity containing ensemble $A$ is channeled to the second cavity containing ensemble $B$, but no light of the latter cavity reaches the first one, cf. Fig.~\ref{fig:uni-quantum setup}. This setup is inherently different from the symmetric configuration in the previous section, and it is unclear if or which synchronization behavior still occurs. What is clear is that the properties of light emitted by ensemble $A$ will be completely unaffected by ensemble $B$ downstream. It is therefore advantageous to assume that the transition frequency of atoms in ensemble $A$ is $\nu$ and the one of atoms in ensemble $B$ is $\nu-\delta$ as indicated in Fig.~\ref{fig:uni-quantum setup}. The cavity frequencies are assumed to be equal to $\nu$, but this assumption is insignificant in the bad cavity limit.

The dynamics in this setup is described by means of a cascaded systems master equation \cite{gardiner_quantum_2004}. In a rotating frame it is given by
\begin{align}
  \dot \rho &= -i \C{  \frac{\Omega}{2} \B{ J_A^+ \a +J_A^- \ad  + J_B^+ \b +J_B^- \bd  }  - \delta J^z_B  }\rho
     \breaksign +   {w \sumsetclap[T=A,B\\i=1\dots N]  \L{\s_{T,i}^+}}    {  + \frac \kappa 2   \C{\ad \b -\bd \a   }\rho } + {\kappa\L{\a+ \b}},  \label{eq:uni-quantum master eq}
\end{align}
The atomic ensembles $A$ and $B$  are coupled to their respective cavity modes $\a$ and $\b$ with single-photon Rabi frequency $\Omega/2$, and are  pumped incoherently at the rate $w $ to their excited states $\ket e$. We dropped already the spontaneous emission into free space and the dephasing, knowing  the enhanced decay   into the cavity modes $\a$ and $\b $ dominate the decay processes, as in the previous section. The  last two terms describe the cascaded, unidirectional coupling and decay of the cavity modes at rate $\kappa$, cf.~\cite{gardiner_quantum_2004}.

As in section \ref{sec:quantum bi} the cavity decay is assumed to be the fastest timescale in the system $\kappa\gg  \Omega,w$, allowing us to adiabatically eliminate the cavity fields which yields here
 \begin{align}
 {\a    }   &\simeq   \frac{\Omega}{i\kappa}    J_A^- , &
 {\b}  &\simeq   \frac{\Omega}{i\kappa}  \B{   J_B^-    -  2   J_A^-    } 
 .
 \label{eq:qu-uni adiabatic elimination}
 \end{align}
The effective master equation for atoms is
\begin{align}
\dot  \rho  &= i \delta     \C{ J^z_B    }{  \rho  }
      +   w \sumsetclap[T,j]  \L[  \rho  ]{\s_{T,i}^+}   \breaksign-\frac \gamma 2    \C{  J_A^+ J_B^- - J_B^+ J_A^-}{ \rho}    +\gamma  \L[ \rho]{J_A^--J_B^-}
      \label{eq:quantum uni master equation atomic}
\end{align}
with $\gamma= \Omega^2/\kappa$.  Comparing this equation to \eqref{eq:qu-bi eliminated} in the previous section  we see that the decay of the two ensembles still happens collectively, despite the relative sign. The additional effective Hamiltonian term describes unidirectional character of the coupling as in Eq.~\eqref{eq:uni-quantum master eq}.

The master equation implies the following equations of motion for the expectation values
\begin{align}
 \partial_t \Mean{\s^z_A} &=-\Mean{\s^z_A}\B{\gamma+w}  -2\gamma    \B{N-1}  \Meansmall{\s^+_A\s^-_A}
    \breaksign-\gamma  +w
 \notag \\
\partial_t \Meansmall{\s^+_A\s^-_A}&=  -  \Meansmall{\s^+_A\s^-_A} \B{\gamma+w -\gamma \Mean{\s^z_A}  \B{ N -2    }   } \breaksign+  \frac \gamma 2  \Mean{\s^z_A} \B{ \Mean{\s^z_A}+1}
 \notag \\
\partial_t \Mean{\s^z_B} &= -  \Mean{\s^z_B}\B{\gamma   +w}-2 \gamma  \B{N-1}    \Meansmall{\s^+_B\s^-_B}
    \breaksign-\gamma +w    + 4 \gamma  N \Re{  \Meansmall{\s^+_A\s^-_B}}
 \notag \\
\partial_t \Meansmall{\s^+_B\s^-_B}&= -   \Meansmall{\s^+_B\s^-_B} \B{\gamma +w-\gamma     \Mean{\s^z_B}  \B{N-2}  } \breaksign   +\frac \gamma 2  \Mean{\s^z_B} \B{  \Mean{\s^z_B}+1}
    \breaksign- 2 \gamma  N  \Mean{\s^z_B} \Re{ \Meansmall{\s^+_A\s^-_B}}
 \notag \\
\partial_t \Meansmall{\s^+_A\s^-_B}&= \Meansmall{\s^+_A\s^-_B} {   \gamma  \B{N-1}  \B{ \Mean{\s^z_A}+ \Mean{\s^z_B}}/2} \breaksign+ \Meansmall{\s^+_A\s^-_B} \B{i    \delta  -\gamma  - w } -\frac \gamma 2  \Mean{\s^z_B}   \Mean{\s^z_A}
\breaksign -\frac \gamma 2  \Mean{\s^z_B} \B{2  \Meansmall{\s^+_A\s^-_A} \B{N-1}+ 1}
\label{eq:uni-quantum expectation values}
,
\end{align}
where we used the symmetry of \eqref{eq:quantum uni master equation atomic} to introduce the abbreviations $\Meansmall{\s^z_A} :=\Meansmall{{\s^z_{A,i}}} $, $\Meansmall{\s^z_B} :=\Meansmall{{\s^z_{B,i}}} $,  $ \Meansmall{{\s^+_A\s^-_A}} =  \Meansmall{{\s^+_{A,i}\s^-_{A,j}}}$, $\Meansmall{{\s^+_B\s^-_B}} = \Meansmall{{\s^+_{B,i}\s^-_{B,j}}}  $ for $i\neq j $, and $ \Meansmall{{\s^+_A\s^-_B}}=\Meansmall{{\s^+_{A,m}\s^-_{B,n}}}$.
 Note that the symmetry between $A$ and $B$ is broken in the cascaded setup. In \eqref{eq:uni-quantum expectation values} we also factorized occurrences of the mean field $\Meansmall{{\s_{A(B)}^z}} $, which we validated using small system numerical solutions of \eqref{eq:uni-quantum master eq}  using  QuTiP \cite{_qutip_2015}.
 The steady state solution  can be obtained by setting all time-derivatives on the left hand sides equal to zero and solving the algebraic equations. The first two  equations involving only ensemble  $A $ can be solved independently of ensemble $B $, as expected in view of the cascaded setup. The remaining equations can be reduced to a polynomial equation  of fourth order, which can be solved exactly and used for analytical results up to leading order in $1/N $. In order to obtain  numerical results it is   easier and faster to solve the  system \eqref{eq:uni-quantum expectation values} numerically  and select the stable solution by linearizing \eqref{eq:uni-quantum expectation values} around each solution.
\begin{figure}[t]\centering
 {\includegraphics[width=\linewidth]{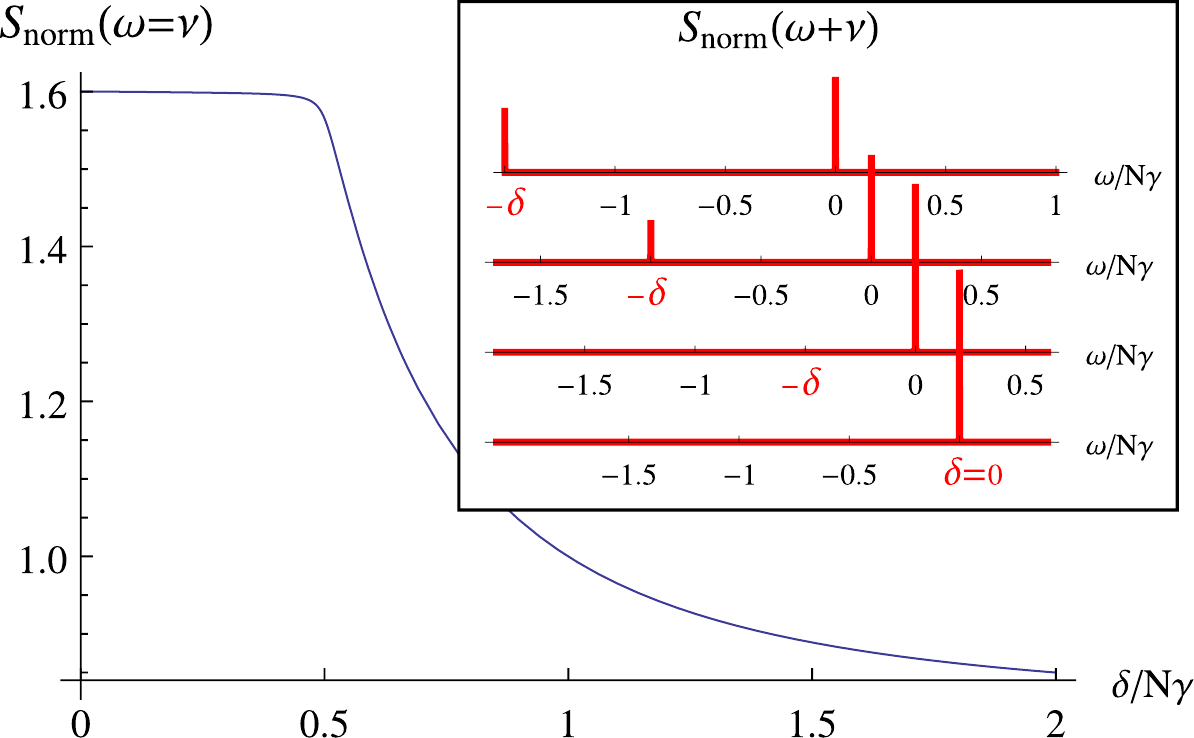}}
  \caption{(Main plot):  $ S_{\text{norm}}(\omega=\nu)$ is the photon flux at the resonance frequency $\nu $ of ensemble $A $. Since ensemble $A $ is independent of $\delta $ any change with $\delta $ comes from ensemble $B $ also  at frequency $\nu $. For decreasing $\delta $ ensemble $B $ radiates stronger on the injected frequency $\nu $ and less at it's resonance frequency $\nu-\delta $. For $\delta<w $ ensemble $B $ radiates dominantly on the injected frequency $\nu $ and decreasing $\delta $ further does not change this, resulting in the plateau of $ S_{\text{norm}}(\omega=\nu)$.
  (Inset):  The normalized spectrum  $ S_{\text{norm}}(\omega)$  for multiple detunings  $\delta/N\gamma =1.5, 1, 0.5, 0 $    shows a suppression of the  peak at $\nu-\delta $ for decreasing detuning $\delta $, while the Lorentz peak at $\nu $ rises   until ensemble $B $ is radiating dominantly at $\nu $. Both plots use for the parameters    $N\gamma=10 \text{ kHz}$ and $w=0.5 N\gamma  $.}
  \label{fig:uni-quantum centralpeakheight}
\end{figure}

 The spectrum  follows again from the Fourier transform of the two-time correlation functions which we calculate using the quantum regression theorem,
 \begin{align}
  \D{\tau} \smallmatrixx{
  \Mean{ \s_A^+(\tau) \s_{T}^-(0)}
  \\
   \Mean{  \s_B^+(\tau)\s_{T}^-(0) }} &=\frac 1 2  \matrixx{X & 0 \\Y& X'}   \smallmatrixx{
  \Mean{ \s_A^+(\tau) \s_{T}^-(0)}
  \\
   \Mean{  \s_B^+(\tau)\s_{T}^-(0) }}
   \label{eq:uni-quantum dynamics of expectation values}
 \end{align}
 with $T=A,B$   and
 \begin{align*}
 X&=     {\gamma}   ( N-1) \Mean{ \s^z_A} - \gamma -w
 \\
 X'&=    {\gamma}     (N-1)\Mean{ \s^z_B} -  \gamma-w - 2 i\delta
 \\
 Y&=    -2N\gamma \Mean{\s^z_B}.
 \end{align*}

The  normalized spectrum \nocite{mandel_optical_1995}  of the field emerging from cavity $\b$ is
\[
S_{\text{norm}}(\omega):=\frac{1}{ {2\pi} I  }     \int \intd {\tau} \Exp{- i \omega \tau}  \Meansmall{ \bd_{\text{out}}(\tau)  \b_{\text{out}}(0)}
\]
which can be evaluated using the input-output relation for cascaded systems  \cite{gardiner_quantum_2004}
\begin{align}
\b_{\text{out}} =\a_{\text{in}} +\sqrt{ \kappa}({ \a+\b}).
\end{align}
and Eq.~\eqref{eq:qu-uni adiabatic elimination} . The normalization factor is $I =  \Meansmall{ \bd_{\text{out}}  \b_{\text{out}}} $. The peaks in $ S_{\text{norm}}(\omega)$  are always  located at the bare transition frequencies $\nu$ and $\nu-\delta $ of of ensemble $A$ and $B$ respectively which does not hint at synchronization effects. Synchronization becomes visible in the regime $\delta<w<N\gamma $ via a change of relative peak heights, as illustrated in Fig.~\ref{fig:uni-quantum centralpeakheight} Inset, which is qualitatively different from the frequency pulling in section \ref{sec:quantum bi}.
 For fixed pumping $w $ in the superradiant regime \cite{meiser_prospects_2009}   $\gamma<w<N\gamma $ we can distinguish different regimes for $\delta $
 \begin{description}
  \item[$\delta\gg w $] Ensembles  $A$ and $B $ radiate only at their own resonance frequency $\nu, \nu-\delta $ respectively with equal intensity.

  \item[$\delta\geq w $] Ensemble $A $ is unaffected by any change in $\delta $ and  radiates at $\nu $, but ensemble $B $ radiates at two frequencies $\nu $ and $\nu-\delta $. This leads to an increasing total intensity at frequency $\nu $, cf. Fig.~\ref{fig:uni-quantum centralpeakheight}.

  \item[$\delta<  w $]  Ensemble $A $ still radiates with the same intensity at frequency $\nu $ and ensemble $B $  now also dominantly radiates at frequency $\nu $, while radiation at its own resonance frequency becomes negligible (for large $N $). Ensemble $B $ is synchronized to ensemble $A $ resulting in a plateau of   $ S_{\text{norm}}(\nu)$, cf. Fig.~\ref{fig:uni-quantum centralpeakheight}.
 \end{description}

\begin{figure}[t]\centering
 {\includegraphics[width=.771\linewidth]{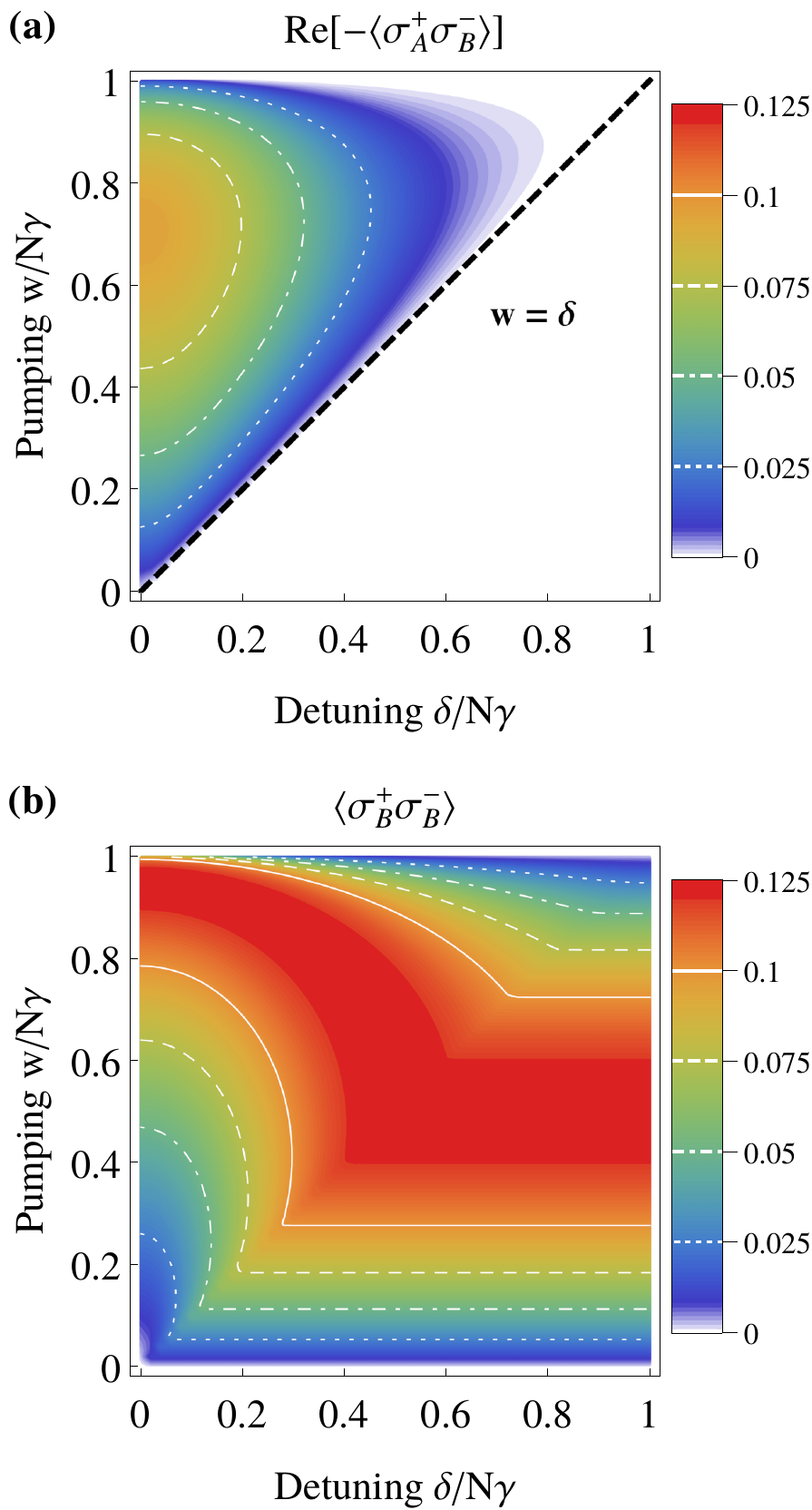}}
  \caption{(a)  Non-vanishing inter-ensemble correlations  ${\Resmall{-\Meansmall{\s^+_A\s^-_B}}   }$ outline the synchronized parameter    regime. The dashed line $w=\delta $ separates the synchronized from the  unsynchronized superradiant regime.
  {(b)} shows the inner-ensemble correlations  $\Meansmall{\s^+_B\s^-_B} $  of the slave ensemble $B $.  For detuning smaller than the incoherent pumping rate $\delta <w$ the slave ensemble $B $  is synchronized to the master ensemble $A $. Here $N\gamma=10^6 \text{Hz} $.}
  \label{fig:uni-quantum contourplotReAB}
\end{figure}

 While the peak value of the normalized spectrum at  frequency $\nu $ shows a plateau  in the synchronized regime $\delta<w $, the integrated unnormalized spectrum, that is the total power output is still increasing for smaller detuning, similar to the finding in section \ref{sec:quantum bi}. In  leading order in  $N^2 $ the total photon flux is given by
\begin{align*}
\Meansmall{\bd_{\text{out}} \b_{\text{out}}} \approx  \gamma   N^2  \B{ ~\sumsetclap[T=A,B]  \Meansmall{ \s_{T,1}^+  \s_{T,2}^-  } +  2 \Re{ -\Mean{ \s_A^+  \s_B^-  }} }.
\end{align*}
The photon flux increases for smaller detuning $\delta $ due to increasing correlations ${\Resmall{-\Meansmall{\s^+_A\s^-_B}}   }$ between the two ensembles  as shown in Fig.~\ref{fig:uni-quantum contourplotReAB}a. In Section~\ref{sec:quantum bi} the synchronized regime stretched out beyond $w=N\gamma $ up to $w=2N\gamma$ for vanishing detunings, cf. Fig.~\ref{fig:bi-quantum contourplotReAB} due to the fact that the two ensembles radiate in this regime as one ensemble containing $2N$ atoms. This is not the case in the cascaded system, see Fig.~\ref{fig:uni-quantum contourplotReAB}. For $w> N\gamma $ ensemble $A$ (containing $N$ atoms) will stop emitting superradiantly and for $w\gg N\gamma $ will radiate chaotic light \cite{meiser_intensity_2010}. The correlations $\Meansmall{\s^+_B\s^-_B} $  shown in Fig.~\ref{fig:uni-quantum contourplotReAB}b indicate that if the first cavity would still radiate superradiantly (e.g. $N $ larger in the first cavity), then the synchronized regime could also stretch beyond $w=N\gamma $.

Analyzing the  Lorentz peaks in the spectrum reveals that the peaks at $\nu$ and $\nu-\delta$  have a width
\begin{align}
\frac{\Gamma_{\nu}}\gamma&=\frac{w}{N\gamma}   +1 ,	\\
\frac{\Gamma_{\nu-\delta}}\gamma&=\begin{cases}
		    O(N)			,& \delta\leq w	\\[1em]
                     \dfrac{w}{N\gamma}   +1 	,& \delta> w
                    \end{cases},
                    \label{eq:qu-uni width of peaks}
\end{align}
which is valid    up to order $1/N $ in the superradiant regime. Most significantly we see that the linewidth at $\nu-\delta $  for $\delta\leq w $ scales with $  N $ and as a result  the peak effectively vanishes for  large $N $.  This shows that ensemble $B $  cannot sustain radiating at its resonance frequency and radiates instead at the frequency of ensemble $A $. The independence of $\Gamma_{\nu} $ of $\delta $  is also significant, since it means that in the synchronized regime  ensemble $B $ is amplifying the input signal without increasing the linewidth.

\section{Synchronization through Classical Channels}

\subsection{Unidirectional Synchronization}
\label{sec:classical uni}

One can ask the question whether the synchronization in Section~\ref{sec:quantum uni} is dominated by   quantum mechanics and requires a quantum channel in between both cavities or whether the same or a similar result can be achieved by synchronizing the two clocks through a classical channel.  Synchronization or locking of the two superradiant laser through a classical channel means that classical information is transmitted between the two systems, rather than quantum states of light as was considered in the previous section.

\begin{figure}[]\centering
{\includegraphics[width=\linewidth]{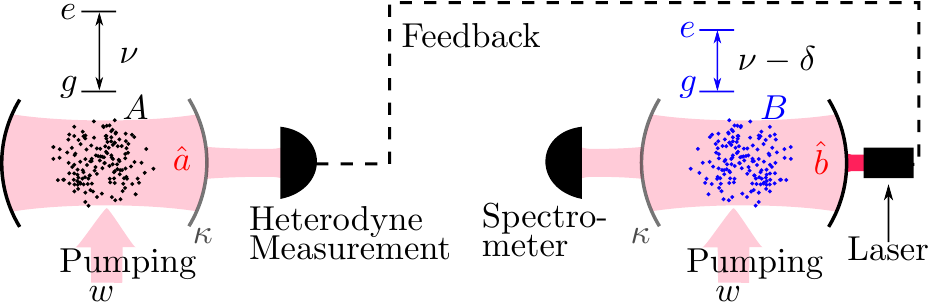}}
  \caption{Two ensembles of two level systems ($A $ and $B $) coupling to the cavity modes $\a,\b $ respectively.  The transition frequencies $\ket g \leftrightarrow \ket e$ of  ensemble $A $ and the cavity frequencies $\a,\b $  are  $\nu $, while   ensemble $B $'s transition frequency  is detuned by $-\delta $.  Atoms are pumped incoherently from $\ket g$ to $\ket e$ via a third fast decaying level (not shown in level scheme) at rate $w $ and decay from $\ket e$ to $\ket g$ predominantly through the cavity.
  The  cavities decay with rate $\kappa $  and the output of cavity $\a $ is measured via an ideal heterodyne detection and then recreated with an ideal laser with a certain gain and fed into  cavity $\b $.  The measurement and feedback  via the laser are a classical simulation of the direct injection in Section~\ref{sec:quantum uni}.
}
  \label{fig:uni-classical setup}
\end{figure}

In this section we are going to answer this questions for a highly idealized classical channel: We will consider phase sensitive measurements (heterodyne detection) of the output field of one laser cavity, transmission of the classical measurement result (the photocurrent), and injection of an appropriate coherent field to the second cavity. Thus, we assume a continuous-time feedback strategy where the measured amplitude and phase of the field of the first cavity is recreated with appropriate feedback gains as a seed for the second cavity as illustrated in Fig.~\ref{fig:uni-classical setup}. This measure and prepare strategy simulates the direct injection of Section~\ref{sec:quantum uni}. Both heterodyne measurement and laser are idealizations adding no technical noise, but will add quantum noise due to the gain of classical information. From a quantum information point of view we have replaced the quantum channel between both cavities by a classical channel and local operations.  We will show  that this introduces a certain level of additional noise due to the measurement, but will not change the synchronization behavior.

This result has to be understood as an upper bound to the quality of classical synchronization achievable through a classical channel. Any real classical procedure will actually perform worse, as it will add technical noise in phase sensitive detection and feedback. This will be especially relevant when attempting to synchronize superradiant lasers exhibiting unprecedentedly low linewidths.

To describe the system we use an unconditional feedback master equation using continuous-time  heterodyne measurements developed in \cite{hofer_time-continuous_2013,wiseman_quantum_1993} 
\begin{align}
\dot \rho &= -i \C{H   }{\rho}   -\frac {i} {4} \C{  \B{\hat F_+ + i \hat F_-}\hat s +h.c.  }{\rho}   \breaksign+\frac 1 2  \L[\rho]{\hat s - i \hat F_+} +\frac 1 2  \L[\rho]{\hat s -  \hat F_-}.    \label{eq:uni-classical feedback master equation start}
\end{align}
The operator $\hat s $ describes the type of measurement being performed which, for the case of a heterodyne detection, is given by $\hat s = \sqrt\kappa \a $. The heterodyne detection provides two photocurrents $I_\pm $ for the phase and the amplitude quadrature which can be used for the feedback operation. We consider Markovian and linear feedback, that is, the photocurrents are each multiplied by suitable gains and, in the case considered here, fed back as a coherent driving field to the second cavity. The feedback due to the two photocurrents $I_\pm $ is described by Hermitian operators  $\hat F_\pm $  which are given by $  \hat F_\pm = g_\pm \b + g_\pm\cc \bd   $ with gain coefficients  $  g_+=-i\sqrt{\kappa}  $ and $ g_-=-\sqrt{\kappa} $. We choose this particular feedback strategy as it reproduces an  unidirectional coupling identical to the one found in Eq.~\eqref{eq:uni-quantum master eq} when inserted to the feedback master equation in \eqref{eq:uni-classical feedback master equation start},
\begin{align*}
\dot \rho &=   -i \C{  \frac{\Omega}{2} \B{ J_A^+ \a +J_A^- \ad  + J_B^+ \b +J_B^- \bd  }  - \delta J^z_B  }\rho
\breaksign+   w \sumsetclap[T=A,B\\i=1\dots N]   \L{\s_{T,i}^+} + \frac \kappa {2}  \C{ \ad \b  -h.c.  }\rho   + \kappa \L{ \a  +  \b}
     \breaksign +     \kappa\L{\b} +\kappa  \L{  \bd}.
\end{align*}
We  added the incoherent atom pumping with rate $w $ and the decay of cavity field $\b $ with rate $\kappa $. The coherent dynamics is given by the atom-cavity interaction at rate $\Omega/2  $, and the detuning of the atomic transitions $-\delta $, as in the previous section. The only difference to \eqref{eq:uni-quantum master eq} are the last to cooling and heating terms indicating additional noise due to the measurement.

\begin{figure}[]\centering
{\includegraphics[width=.771\linewidth]{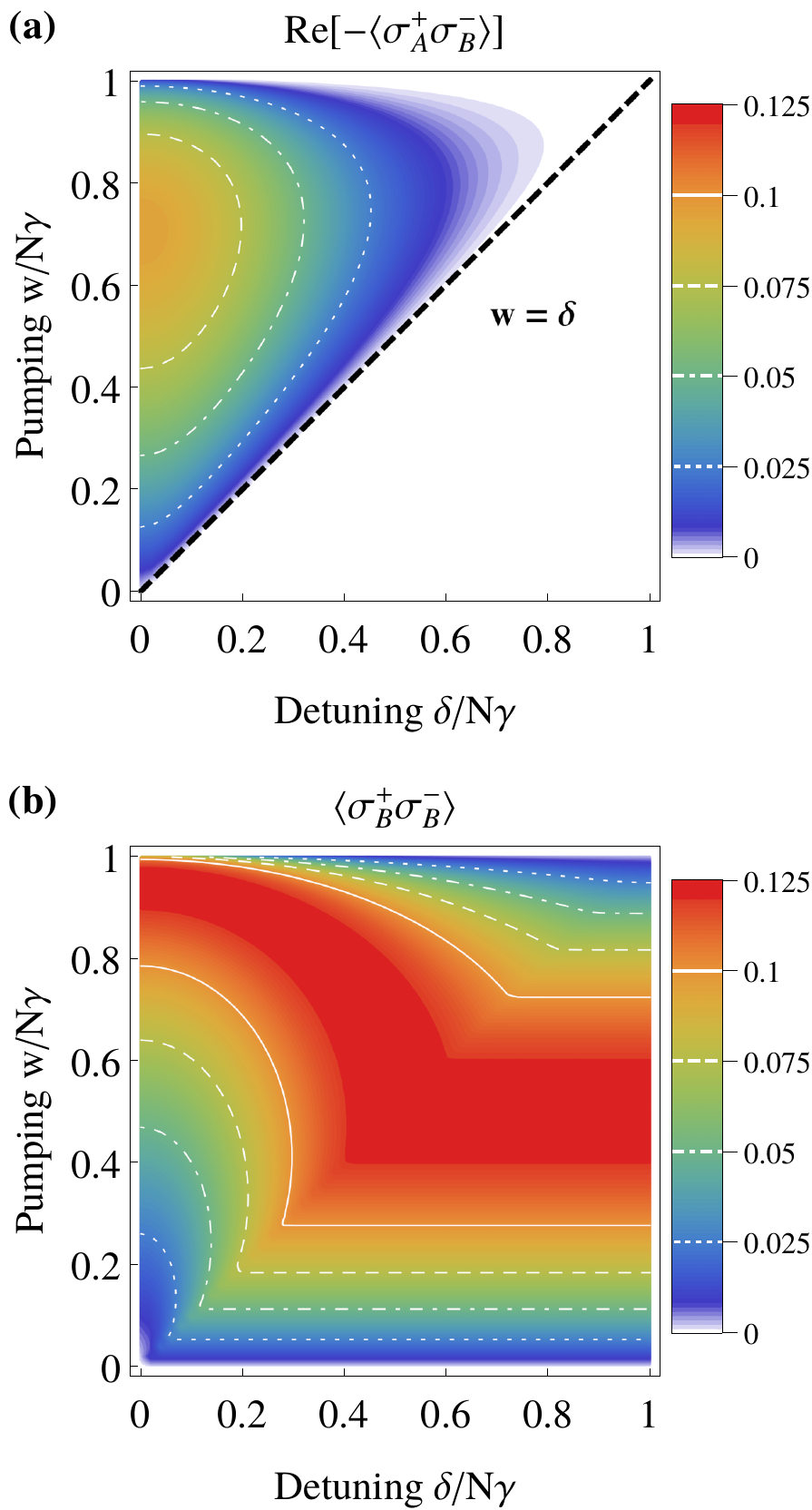}}
  \caption{(a)  Non-vanishing inter-ensemble correlations  ${\Resmall{-\Meansmall{\s^+_A\s^-_B}}   }$ outline the synchronized parameter    regime. The dashed line $w=\delta $ separates the synchronized from the  unsynchronized superradiant regime and is identical to Sec.~\ref{sec:quantum uni}.
  {(b)} shows the inner-ensemble correlations  $\Meansmall{\s^+_B\s^-_B} $  of the slave ensemble $B $.    Here $N\gamma=10^6 \text{Hz} $.  }
  \label{fig:uni-classical contourplotReAB}
\end{figure}

The cavity fields can again be adiabatically eliminated  considering the subtlety that cavity $\b $ is now driven with rate $\kappa $  by the Lindblad terms to a thermal state with $1 $ mean photon. The adiabatic elimination translates the decays of the cavity modes  to a collective decay of the atoms  at rate  ${\gamma}= {\Omega^2}/{\kappa} $
\begin{align*}
\dot  \rho  &=   i \delta     \C{ J^z_B    }\rho  +\gamma \L{J_A^- -J_B^-}        +\frac \gamma 2  \C{J_B^+ J_A^- - h.c.}\rho
        \breaksign +   w \sumsetclap[T=A,B\\i=1\dots N]  \L{\s_{T,i}^+}   + \gamma         \L{J_B^-}   +\gamma\L{J_B^+}
.
\end{align*}
The corresponding dynamics of the  expectation values is
\begin{align}
 \partial_t \Mean{\s^z_A} &=-\Mean{\s^z_A}\B{\gamma+w}  -2\gamma    \B{N-1}  \Meansmall{\s^+_A\s^-_A}
    \breaksign-\gamma  +w
 \notag \\
\partial_t \Meansmall{\s^+_A\s^-_A}&=  -  \Meansmall{\s^+_A\s^-_A} \B{\gamma+w -\gamma \Mean{\s^z_A}  \B{ N -2    }   } \breaksign+  \frac \gamma 2  \Mean{\s^z_A} \B{ \Mean{\s^z_A}+1}
 \notag \\
\partial_t \Mean{\s^z_B} &= -  \Mean{\s^z_B}\B{\gamma {u}  +w}-2 \gamma  \B{N-1}    \Meansmall{\s^+_B\s^-_B}
    \breaksign-\gamma +w    + 4 \gamma  N \Re{  \Meansmall{\s^+_A\s^-_B}}
 \notag \\
\partial_t \Meansmall{\s^+_B\s^-_B}&= -   \Meansmall{\s^+_B\s^-_B} \B{{u}\gamma +w-\gamma     \Mean{\s^z_B}  \B{N-2}  } \breaksign   +\frac \gamma 2  \Mean{\s^z_B} \B{ {u} \Mean{\s^z_B}+1}
    \breaksign- 2 \gamma  N  \Mean{\s^z_B} \Re{ \Meansmall{\s^+_A\s^-_B}}
 \notag \\
\partial_t \Meansmall{\s^+_A\s^-_B}&= \Meansmall{\s^+_A\s^-_B} {   \gamma  \B{N-1}  \B{ \Mean{\s^z_A}+ \Mean{\s^z_B}}/2} \breaksign+ \Meansmall{\s^+_A\s^-_B} \B{i    \delta  -{v}\gamma  - w } -\frac \gamma 2  \Mean{\s^z_B}   \Mean{\s^z_A}
\breaksign -\frac \gamma 2  \Mean{\s^z_B} \B{2  \Meansmall{\s^+_A\s^-_A} \B{N-1}+ 1}
\label{eq:uni-classical expectation values}
,
\end{align}
where ${u}=3$ and ${v}=2$. This is almost identical to the dynamics found for the cascaded system considered in the previous section, Eqs.~\eqref{eq:uni-quantum expectation values} and \eqref{eq:uni-quantum dynamics of expectation values}, which are identical the set of equations in \eqref{eq:uni-classical expectation values} when the parameters $u$ and $v$ are set to ${u}=1$ and $ {v}=1  $. Importantly,  $u $ and  $v $ never occur multiplied with  $N,w, \delta $  and therefore  do not contribute  significantly to the dynamics in the limit of large $N $. This is also visible in the steady state results in Figs. \ref{fig:uni-classical contourplotReAB} and showing no visible difference to Fig.~\ref{fig:uni-quantum contourplotReAB}.

To evaluate if ensemble $B $ synchronizes with ensemble $A $, just like in Section~\ref{sec:quantum uni}, we extract from the two-time correlation functions the components $ \Exp{- {\Gamma_{\nu}}   \tau/ 2 }$ and $ \Exp{-  \B{  {\Gamma_{\nu-\delta}}/ 2  +  i \delta    }\tau }$. Using the solutions for $\Meansmall{\s^z_A} $ and $\Meansmall{\s^z_B} $, which are  are identical to Section~\ref{sec:quantum uni} up to leading order in $1/N $, we calculate the width of these Lorentzian peaks, giving in the superradiant regime
\begin{align*}
\frac{\Gamma_{\nu}}\gamma&=\frac{w}{N\gamma}   +1 ,	&
\frac{\Gamma_{\nu-\delta}}\gamma&=\begin{cases}
		    O(N)			,& \delta\leq w	\\
                     \dfrac{w}{N\gamma}   +3 	,& \delta> w
                    \end{cases}
.
\end{align*}
Just as in Section~\ref{sec:quantum uni}  we  see that the peak at $\nu-\delta $  for $\delta\leq w $ gets extremely broad  for  large $N $ and thus effectively vanishes. Again this means that the   resonance frequency of ensemble $B $  is suppressed and ensemble $B $  synchronizes to the frequency of ensemble $A $.  Remarkable is that even though there is now a classical channel between both cavities, ensemble $B $ amplifies the input signal  in the synchronized regime  without increasing the linewidth $\Gamma_{\nu} $.  In the unsynchronized regime $\delta>w $    the linewidth   $ \Gamma_{\nu-\delta} $  is larger than in the quantum coupled setup \eqref{eq:qu-uni width of peaks}. Due to the chosen gain in the feedback operators  $  \hat F_\pm = g_\pm \b + g_\pm\cc \bd   $  the output spectrum of cavity $\b $ has now a larger Lorentz peak at  $\nu $ than at $\nu-\delta $ for large detuning $\delta\gg w $. This stronger feedback gain  is necessary to simulate the same amplitude of cavity field $\a $ being injected into cavity  $\b $ as in Sec.~\ref{sec:quantum uni}.

From  the dynamics of the expectation values \eqref{eq:uni-classical expectation values}  and from the correlation functions in the steady state Fig.~\ref{fig:uni-classical contourplotReAB} we see that there is no significant difference in the synchronization between the quantum and the classically coupled setups considered in the previous and this section, respectively. This holds in the limit of large $N$, that is far above threshold of the superradiant laser where the emitted field is essentially classical.  However, it is important to remember that our analysis is based on an ideal heterodyne detection and feedback operations, and that any realistic classical synchronization will perform worse.

\subsection{Bidirectional Synchronization}
\label{sec:classical bi}

\begin{figure}[t]\centering
{\includegraphics[width=\linewidth]{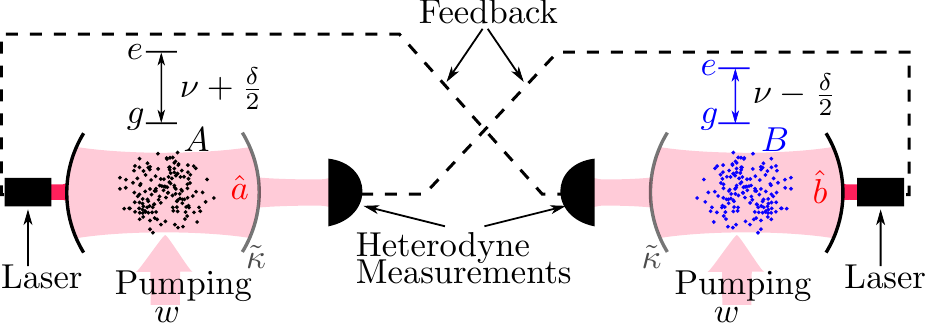}}
  \caption{ Two ensembles of two level systems ($A $ and $B $) coupling to the cavity modes $\a,\b $ respectively.  The frequencies of the transitions $\ket g\leftrightarrow\ket e$ are detuned by $\pm\delta/2$ from the cavity resonance at frequency $\nu$ for ensemble $A$ and $B$ respectively.   Atoms are pumped incoherently from $\ket g$ to $\ket e$ via a third fast decaying level (not shown in level scheme) at rate $w $ and decay from $\ket e$ to $\ket g$ predominantly through the cavity.   The  cavities decay with rate $\Kappa $  and the output of both cavities  is measured via an ideal heterodyne detection and then recreated with an ideal laser with a certain gain and fed into  the opposite cavity.  The measurements and feedbacks via the lasers are symmetric such that this simulates the coupling to the same cavity mode as in Sec.~\ref{sec:quantum bi}.  }
  \label{fig:bi-classical setup}
\end{figure}

In view of the results of the previous section it is worthwhile considering the question whether the synchronization in Section~\ref{sec:quantum bi} was dependent on the coupling to the same quantum mechanical cavity mode, or if this synchronization also occurs when we replace this quantum coupling with a classical, bidirectional coupling. In order to address this question we consider the setup in Fig.~\ref{fig:bi-classical setup}. Both cavity fields decay with rate $\Kappa $ and are measured with ideal heterodyne measurements. The measurement results are then used by an ideal lasers to recreate the measured coherent state with a certain gain, giving rise to a symmetric coupling between both cavities using classical channels.
Just like in the previous section are the heterodyne measurements and lasers are idealizations adding no technical noise and the continuous-time feedback is  instantaneous -- i.e. Markovian.
This setup is a strategy to simulate the coupling to the same cavity mode  in Section~\ref{sec:quantum bi} with a classical  (but not necessarily technical feasible) bi-directional coupling.
In this section we will give a brief overview over our analysis and its results and refer to the Appendix
for a complete derivation.

To describe this system we use the same unconditional feedback master equation \eqref{eq:uni-classical feedback master equation start} twice. Once with the measurement operator  $\hat s_\a = \sqrt\Kappa \a $ and feedback operators $  \hat F^\b_\pm = g_\pm \b + g_\pm\cc \bd   $ acting on field $\b $, and then with  the  measurement operator $\hat s_\b = \sqrt\Kappa \b $  and feedback operator $  \hat F^\a_\pm = g_\pm \a + g_\pm\cc \ad   $ acting on field $\a $, where $g_+:=g_-/i $.  Without loss of generality we can introduce the feedback strength $\xi $ with $g_- := -\xi \sqrt\Kappa $ and restrict  the feedback strength to $\xi \in [0,1) $, such that the resulting equations form a stable system for the cavity fields. If $\xi $ would be allowed to be equal to unity or larger, the measurement \& feedback would increase the amplitude of the cavity fields and there would be no steady state with   finite amplitudes.
We can proceed with adiabatically  eliminating the cavity fields, which  gives the master equation for the atoms only
\begin{align}
  \dot \rho   &=   \frac{\delta}{2i}     \C{   J^z_A -J^z_B      }\rho
      +   \sumsetclap[T\in\Curly{A,B}\\i\in \Curly{1..N}] w \L{\s_{T,i}^+}  +      \sumsetclap[s=\pm] \frac{\Omega^2}{2 \kappa_s } \times
   \breaksign{\times} \bracketsize{2}( \B{1+\bar n_s}  \L[]{J_A^- -s J_B^-}+\bar n_s  \L[]{J_A^+  -s J_B^+}     \bracketsize{2})    \rho
    \label{eq:bi-classical ME after adiabatic elimination}
,
\end{align}
where $\kappa_\pm := \Kappa   \B{1\pm\xi} $ and  $\bar n_\pm :=  {\xi^2}/\B{4 (1\pm\xi)}  $.

For $\xi=0 $ the second Lindblad terms drop out and the first Lindblad terms can be transformed to show independent decay for both ensembles. For $\xi\neq 0$ the  the Lindblad terms cannot be separated for both ensembles and for increasing $\xi $ both ensembles couple more and more strongly.
Comparing the dynamics of $\Meansmall{\s_A^+} $ with the completely uncoupled case and the completely coupled case in Sec.~\ref{sec:quantum bi} we choose the cavity decays $\Kappa $ dependent on the feedback strength $ \xi$ such that both cases are simulated best:
  \begin{align}
  \Kappa(\xi)&:= {\frac{\kappa }{(1-\xi)(1+\xi)}}
  \label{eq:parameterization of Kappa}
.
  \end{align}

\begin{figure}[t]\centering
{\includegraphics[width=.771\linewidth]{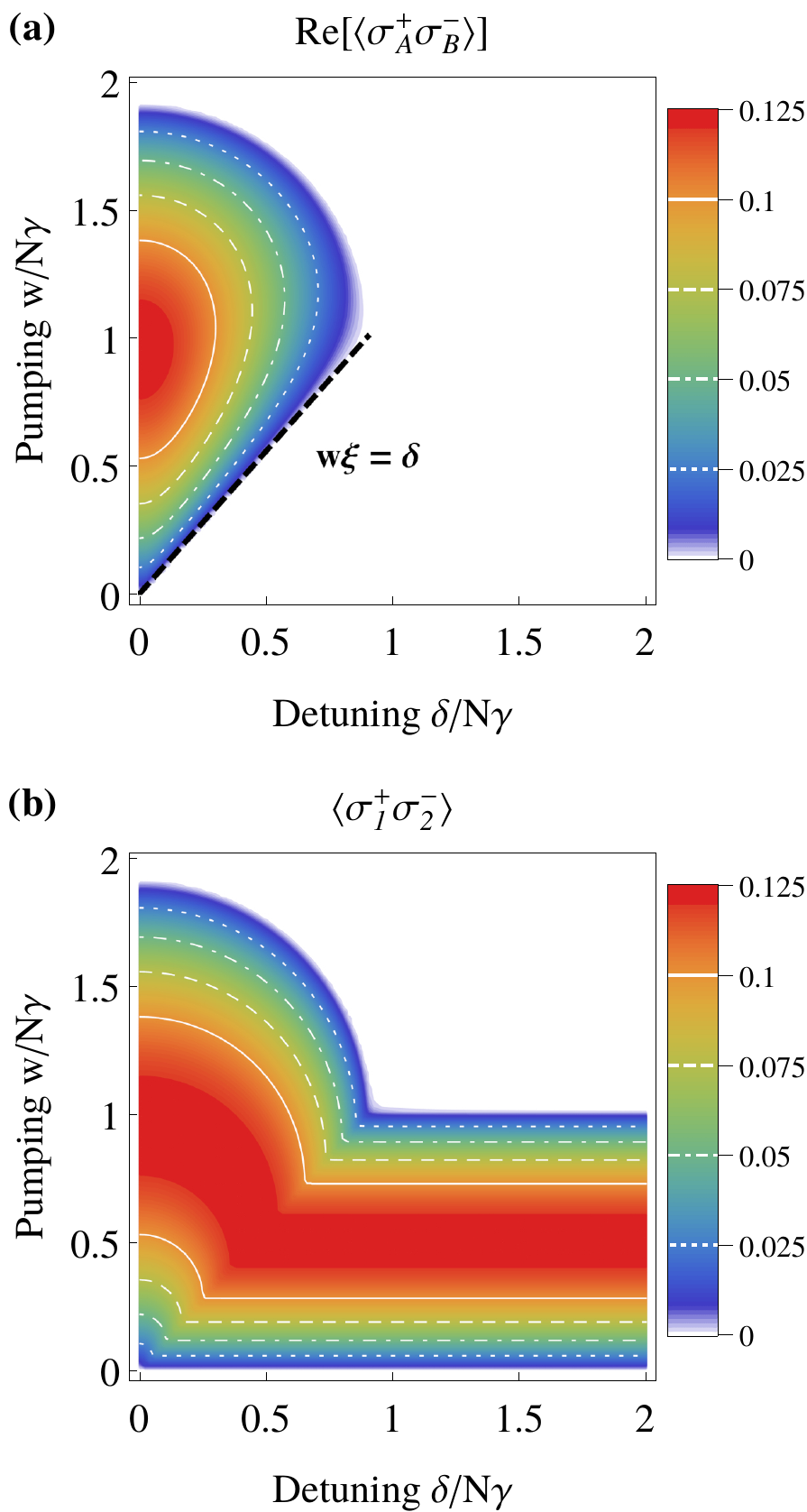}}
  \caption{
  (a) Non-vanishing inter-ensemble correlations  $\Resmall{\Meansmall{\s_A^+ \s_B^-}}   $  outline the synchronized parameter
   regime. This regime is reduced   compared to Fig.~\ref{fig:bi-quantum contourplotReAB}, and the dashed line $w\xi=\delta $ separates the synchronized from the  unsynchronized superradiant regime.
   {(b)} shows the inner-ensemble correlations  $\Meansmall{ \s_1^+  \s_2^-}   $ equal for both ensembles.    For detuning smaller than the  incoherent pumping rate times the feedback strength $\delta <w\xi$ both ensembles  are synchronized and the critical pumping rate is moved from $w=N\gamma $ for $\delta> w\xi $  to $w=(1+\xi)N\gamma $ for $\delta=0 $. Both plots use  $N\gamma=10^6 \text{Hz} $.
   }
  \label{fig:bi-classical contourplotReAB}
\end{figure}

\begin{figure}[t]\centering
{\includegraphics[width=\linewidth]{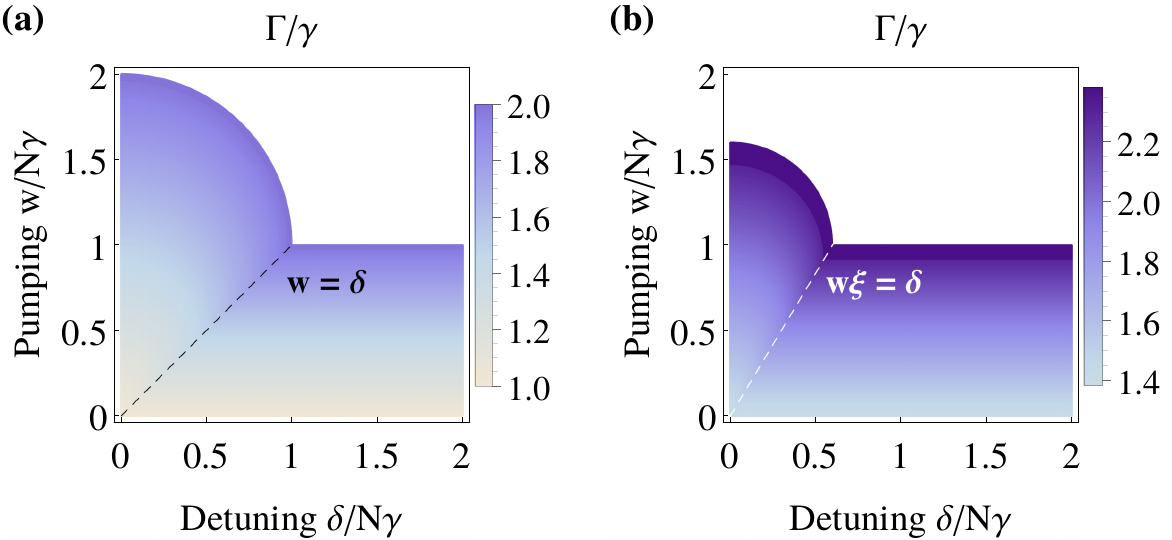}}
 \caption{The dimensionless linewidth $\Gamma /\gamma$    for quantum (a) Eq.~\eqref{eq:quantum bi linewidth} and classical (b) Eq.~\eqref{eq:classical bi linewidth} coupling with $\xi=0.6 $ for leading  order in $1/N $.       The linewidth for classical coupling (b)    is always larger than the quantum coupling, due to noise term $\zeta $ increasing with coupling strength $\xi $.  In the regime far above a critical pumping    the atoms radiate chaotically \cite{meiser_intensity_2010} with a  linewidth scaling with $O(N)$, which is not plotted here and  typically many orders of magnitude larger than the linewidth in the superradiant regime.}
 \label{fig:bi-classical peak wdth comparison}
\end{figure}

\begin{figure}[t]\centering
{\includegraphics[width=.95\linewidth]{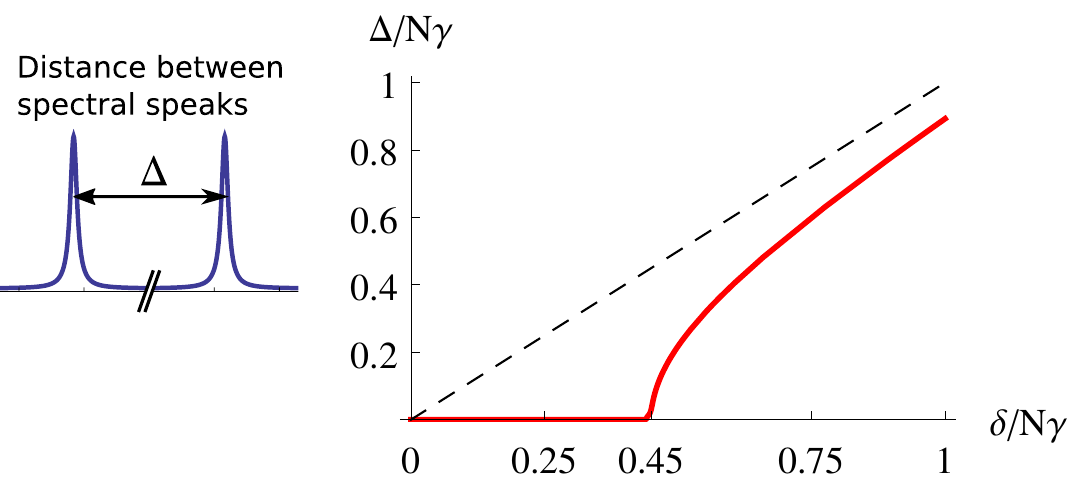}}
 \caption{Pole distance $\Delta $ of the output spectrum over the detuning $\delta $ with the critical detuning at $w \xi $, and the parameters  $N\gamma=10^6 \text{Hz} $, $ w=0.5 N\gamma $, and $\xi=0.9 $. The dashed line is $\Delta=\delta $.}
 \label{fig:bi-classical pole distance}
\end{figure}

From   master equation \eqref{eq:bi-classical ME after adiabatic elimination} with the parameterization \eqref{eq:parameterization of Kappa} we can calculate the dynamics of the expectation  values
\begin{align}
  \partial_t \Mean{\s^z} &=    w\B{  1-   \Mean{\s^z}} -  \gamma -  \Mean{\s^z} \gamma \zeta
   \breaksign-  2\gamma    \B{  \Mean{\s^+_1\s^-_2} (N-1) + \xi N \Re{\Mean{\s^+_A\s^-_B}}   }
  \notag\\
  \partial_t \Mean{\s^+_1\s^-_2} &=  \Mean{\s^+_1\s^-_2}\B{- w  +  \gamma \B{N-2} \Mean{\s^z}-\gamma\zeta   }
  \breaksign +\frac {\gamma}{2  } \Mean{\s^z} \B{1  +\zeta  \Mean{\s^z}+2N \xi\Re{\Mean{\s^+_A\s^-_B}} }
  \notag\\
  \partial_t \Mean{\s^+_A\s^-_B} &=\Mean{\s^+_A\s^-_B} \B{  \gamma  \B{N-1}  \Mean{\s^z}  -\gamma \zeta  -  w+i \delta }
  \breaksign+  \frac \gamma 2  \xi \Mean{\s^z} \B{  2 \Mean{\s^z} \zeta\B{\xi^4 -\xi^2+2 }^{-1}  + 1}
  \breaksign+    \gamma   \xi \Mean{\s^z}    \Mean{\s^+_1\s^-_2} \B{N-1}
  \label{eq:classical bi dynamics of expectation values}
\end{align}
with   $\zeta:= \B{\xi ^4-\xi ^2+2}/\B{2\B{1-\xi^2}}   $ and   factorized $\Meansmall{\s^z} $ from all occurring correlation functions, giving  a closed  system of equations. We could use the same short notation for the expectation values  as in Sec.~\ref{sec:quantum bi}, since the effective coupling in \eqref{eq:bi-classical ME after adiabatic elimination} is symmetric for both ensembles and even recover the equations of \cite{xu_synchronization_2014}, when disregarding the relation between  $\zeta $ and  $\xi $ and setting $\zeta=\xi=1 $. The steady state can now simply be calculated setting all time-derivatives equal  to zero. These algebraic equations can be solved  numerically or analytically, while filtering out the stable solution.
Fig.~\ref{fig:bi-classical contourplotReAB} show the expectation values responsible for inter- and inner- ensemble correlations in the steady state and they are very similar to Fig.~\ref{fig:bi-quantum contourplotReAB} in Section~\ref{sec:quantum bi}. The only difference is a by $\xi $ reduced synchronization regime, visible in non-vanishing inter-ensemble correlations $  \Meansmall{\s_A^+   \s_B^-}  $. The synchronized regime for leading order in $1/N $  is bounded  by  $w\xi=\delta $ and the quarter circle $ (w-N\gamma)^2 +\delta^2 =(\xi N\gamma)^2 $  (see Fig.~\ref{fig:bi-classical contourplotReAB}a).

Analog to Section~\ref{sec:quantum bi} we can extract the half-width $\Gamma/2 $ of the Lorentz peaks in the spectrum from the two-time correlation functions and use the analytical solution for $\Meansmall{\s^z} $ (see \eqref{eq:Aclassical bi solution z})  up to leading order in $1/N $ to derive
\begin{align}
\Gamma/\gamma &=   \zeta+
 \begin{cases}
                    \frac{w-\sqrt{w^2\xi^2-\delta^2\B{1-\xi^2}}}{N\gamma\B{1-\xi^2}}       ,&   0\leq \delta<w \xi	\\
                 \frac{w}{N\gamma}        ,&   \delta\geq w \xi 	
               \end{cases}
               \label{eq:classical bi linewidth}
,
\end{align}
which is valid in   the superradiant regime. This   linewidth  is plotted in Fig.~\ref{fig:bi-classical peak wdth comparison}b and compared with  \eqref{eq:quantum bi linewidth} from Section~\ref{sec:quantum bi} plotted in Fig.~\ref{fig:bi-classical peak wdth comparison}a.
The linewidth using the classical coupling  is larger than the linewidth using the quantum coupled setup, due to the measurement induced noise term $\zeta $.
This noise term $\zeta  $ also prevents one to take the limit $\xi\to 1 $ to approach the same  synchronization regime as in Section~\ref{sec:quantum bi}, since $\zeta  $ diverges in this limit.

%
The striking feature of the setup  in Section~\ref{sec:quantum bi} was clear synchronization visible in  the  distance  of the Lorentzian peaks $\Delta $ plotted over the bare detuning $\delta $, which  is also reproduced here  with a smaller critical detuning $ w\xi $ (see Fig.~\ref{fig:bi-classical pole distance}).

The results presented here show that the  synchronization of superradiant lasers \cite{xu_synchronization_2014,weiner_phase_2015}   is not dominated by  quantum effects, but a classical synchronization of quantum systems.
However this classical coupling setup has a reduced synchronization regime and an increased linewidth, even with the ideal measurements and lasers assumed for the feedback.  For any experimental realization the lasers for the feedback would need an even lower linewidth than the superradiant lasers. This setup is therefore of more theoretical interest to help defining the border between  synchronization of quantum systems using classical channels  and  quantum systems using quantum channels.

\section{Discussion}
\label{sec:discussion}

We discussed if and how synchronization occurs in a cascaded setup of master \& slave superradiant atomic ensembles, or active atomic clocks. Additionally we simulated the symmetric coupling and the cascaded coupling with idealized classical coupling channels.

The cascaded setup in Section~\ref{sec:quantum uni}   shows   synchronization of the slave ensemble  to the injected frequency. The main difference to Section~\ref{sec:quantum bi}    is that the synchronization is not apparent in the  distance $\Delta $ of the Lorentz peaks, but in the  Lorentz peak heights. In the synchronized regime the slave ensemble radiates only at the injected frequency, while it's Lorentz peak at the resonance frequency effectively vanishes.

In Section~\ref{sec:classical uni} we replaced the direct   injection of the light with a measurement and feedback, introducing a classical channel in between both cavities. The resulting steady state equations reveal only   minor changes, which do not scale with the system size, resulting in basically identical steady state results and the same synchronization far above laser threshold.
%

\begin{acknowledgements}
 This work was funded by the Deutsche Forschungsgemeinschaft (DFG) in the research project  RTG 1991  and through CRC 1227 (DQ-mat), project A05.  We thank Hashem Zoubi and  Jonas Lammers for fruitful discussions.
\end{acknowledgements}

\appendix*
\section{Complete Derivation for the Bidirectional Synchronization using a Classical Channel}
\label{sec: Appendix Bidirectional Synchronization using a Classical Channel}

In this section we give a complete derivation of the results presented in Section~\ref{sec:classical bi}. The system (see Fig.~\ref{fig:bi-classical setup})  is comprised out of two one-sided cavities, with  decay   rate $\Kappa $ and are measured with ideal heterodyne measurements. The measurement results are then used by   ideal lasers to recreate the measured coherent state with a certain gain. This can then be injected through the fully reflecting mirror  by considering the limit of vanishing transmission and infinitely large laser gain resulting in a constant amplitude of the injected signal. Since measurement and feedback are symmetric, they realize a symmetric classical coupling channel between both cavities.
To describe the system we use the   unconditional feedback master equation \eqref{eq:uni-classical feedback master equation start} twice. Once with the measurement operator  $\hat s_\a = \sqrt\Kappa \a $ with cavity decay rate $\Kappa $ and feedback operators $  \hat F^\b_\pm = g_\pm \b + g_\pm\cc \bd   $, and then with  the  measurement operator $\hat s_\b = \sqrt\Kappa \b $  and feedback operator $  \hat F^\a_\pm = g_\pm \a + g_\pm\cc \ad   $, where $g_+:=g_-/i $.  Without loss of generality we can define the feedback strength as  $g_- := -\xi \sqrt\Kappa $ with $\xi\in [0,\oo) $ giving  the master equation in a rotating frame:
\begin{align*}
\dot \rho &=      -i \C{  \frac{\Omega}{2} \B{ J_A^+ \a
+ J_B^+ \b
+h.c.
}  + \frac \delta 2 \B{  J^z_A -J^z_B     }   }\rho
\breaksign+   w \sumsetclap[T\in\Curly{A,B}\\i\in \Curly{1..N}]   \L{\s_{T,i}^+}
  + \Kappa \L{   \a +\xi  \b } +  \xi^2 \Kappa  \L{  \bd }  				
 \breaksign + \Kappa \L{   \b +\xi  \a } + \xi^2 \Kappa  \L{  \ad }  				
.
\end{align*}

\subsubsection{Stability}
First it is important to recognize the stability regime  of this feedback for the parameters $\Kappa$ and $\xi $. One might think of the case where the feedback is  effectively larger than the measurement result, giving in a net amplification and diverging amplitudes of the cavity fields. The dynamics of the expectation values
\begin{align}
\D{t} \matrixx{\Mean\a \\ \Meansmall\b} &= -  \frac \Kappa 2   \matrixx{1     & \xi   \\ \xi   &  1  }  \matrixx{\Mean\a \\ \Meansmall\b}
 \label{eq:stability matrix}
\end{align}
including  only the fields is a stable system, if and only if  all eigenvalues $-\frac{1}{2}\Kappa  \B{1\pm   \xi }$  are negative.
This gives the stability condition    $ \xi<1$. Furthermore  \eqref{eq:stability matrix} shows that for $ \xi=0 $  the fields $\a,\b $ are completely decoupled, while for $\xi\lesssim 1$ the fields couple strongly.

\subsubsection{Adiabatic Elimination}
Using the  reparameterization    $ \c_+:=  ({\b-\a})/\sqrt 2 $ and $\c_-:=({ \b+\a})/\sqrt 2 $ the Lindblad operators decouple and  drive the fields $\c_\pm $ into a thermal product state. Following the adiabatic elimination in this reparameterization  we get the master equation for the atoms only
\begin{align}
  \dot \rho   &=   \frac{\delta}{2i}     \C{   J^z_A -J^z_B      }\rho
      +   \sumsetclap[T\in\Curly{A,B}\\i\in \Curly{1..N}] w \L{\s_{T,i}^+}  +      \sumsetclap[s=\pm] \frac{\Omega^2}{2 \kappa_s } \times
   \breaksign{\times} \bracketsize{2}( \B{1+\bar n_s}  \L[]{J_A^- -s J_B^-}+\bar n_s  \L[]{J_A^+  -s J_B^+}     \bracketsize{2})    \rho
    \label{eq:Abi-classical ME after adiabatic elimination}
,
\end{align}
where $\kappa_\pm := \Kappa   \B{1\pm\xi} $ and  $\bar n_\pm :=  {\xi^2}/\B{4 (1\pm\xi)}  $.

\subsubsection{Coupling Parameterization}
Two free parameters $\Kappa $ and $\xi $ remain in \eqref{eq:Abi-classical ME after adiabatic elimination}. We would like however to have one free parameter tuning between decoupled cavities and strongly coupled cavities simulating the  setup in section \ref{sec:quantum bi}.
We can fix the remaining free parameter $\Kappa $ using the  dynamics of the expectation value
\begin{align}
\D{t} \Mean{\s_A^+} &= \Mean{\s^+_A}   \sumsetclap[s=\pm]   \frac {\Omega ^2} {4 \kappa_s }      \B{\B{N-1}  \Mean{\s^z}-1 -2   \bar n_s}
 \breaksign+  \Mean{\s^+_A}   \frac{i \delta-w}{2   }      -   \Mean{\s^+_B}   \Mean{\s^z}  N \sumsetclap[s=\pm]     \frac{s\Omega ^2} {4   \kappa_s}
 \label{eq:bi-classical dynamics of s^+}
 ,
\end{align}
where we factorized  the mean field $\Meansmall{\s^z} $, and compare it to the dynamics for uncoupled cavities  and strongly coupled cavities as in Section~\ref{sec:quantum bi}.

\begin{description}
 \item[Uncoupled, $\xi=0$]   \eqref{eq:bi-classical dynamics of s^+}  simplifies to:
  \begin{align*}
  \D{t} \Mean{\s_A^+} &=  \Mean{\s^+_A}   \B{  \frac {\Omega ^2} {\Kappa } \B{  \B{N-1}  \Mean{\s_A^z}   -1}  -w+i \delta        }/2	
  \end{align*}
  This can be compared to the top left matrix element of \eqref{eq:uni-quantum dynamics of expectation values} in Sec.~\ref{sec:quantum uni}, since   \eqref{eq:uni-quantum dynamics of expectation values} is was derived using the quantum regression theorem and holds identically also for the dynamics of $  (\Meansmall{\s_A^+} ,\Meansmall{\s_B^+} )^T  $.  Considering a shifted  frequency detuning  $\pm \delta/2 $   restricts $\Kappa(\xi=0) =  \kappa  $.

 \item[Strongly coupled, $\xi\to 1 $] The coefficients in \eqref{eq:bi-classical dynamics of s^+} should be identical to the top left  matrix element of \eqref{eq:bi-quantum dynamics of expectation values} in Sec.~\ref{sec:quantum bi}  for all variables scaling with the system size $N $. This gives the restriction  for $\xi\to 1 $:
\begin{align*}
\Kappa(\xi) =   \frac{ \kappa}{2(1-\xi)}
  .
\end{align*}
For $\xi $ very close to unity  the term not scaling with system size,  $\Meansmall{\s^+_A}   \sum_{s=\pm}   \frac {\Omega ^2} {4 \kappa_s }      \B{- 1 -2   \bar n_s}   $, diverges and  becomes dominant. This  noise term    grows \mbox{$\propto (1-\xi)^{-1} $} when $\xi $ approaches the stability border and is negligible   for completely decoupled systems and large $N $.

 \item[Medium coupling] Satisfying both extreme cases discussed before we   can choose  in between:
  \begin{align}
  \Kappa(\xi)&:= {\frac{\kappa }{(1-\xi)(1+\xi)}}
  \label{eq:Aparameterization of Kappa}
.
  \end{align}
\end{description}

\subsubsection{Steady State}
From   master equation \eqref{eq:Abi-classical ME after adiabatic elimination} with the parameterization \eqref{eq:Aparameterization of Kappa} we can calculate the dynamics of the expectation values of $  \Meansmall{\s^z},  \Meansmall{ \s^+_1\s^-_2 },  \Meansmall{\s^+_A\s^-_B} $
\begin{align}
  \partial_t \Mean{\s^z} &=    w\B{  1-   \Mean{\s^z}} -  \gamma -  \Mean{\s^z} \gamma \zeta
   \breaksign-  2\gamma    \B{  \Mean{\s^+_1\s^-_2} (N-1) + \xi N \Re{\Mean{\s^+_A\s^-_B}}   }
  \notag\\
  \partial_t \Mean{\s^+_1\s^-_2} &=  \Mean{\s^+_1\s^-_2}\B{- w  +  \gamma \B{N-2} \Mean{\s^z}-\gamma\zeta   }
  \breaksign +\frac {\gamma}{2  } \Mean{\s^z} \B{1  +\zeta  \Mean{\s^z}+2N \xi\Re{\Mean{\s^+_A\s^-_B}} }
  \notag\\
  \partial_t \Mean{\s^+_A\s^-_B} &=\Mean{\s^+_A\s^-_B} \B{  \gamma  \B{N-1}  \Mean{\s^z}  -\gamma \zeta  -  w+i \delta }
  \breaksign+  \frac \gamma 2  \xi \Mean{\s^z} \B{  2 \Mean{\s^z} \zeta\B{\xi^4 -\xi^2+2 }^{-1}  + 1}
  \breaksign+    \gamma   \xi \Mean{\s^z}    \Mean{\s^+_1\s^-_2} \B{N-1}
  \label{eq:Aclassical bi dynamics of expectation values}
\end{align}
with   $\zeta:= \B{\xi ^4-\xi ^2+2}/\B{2\B{1-\xi^2}}   $ and   factorized $\Meansmall{\s^z} $ from all occurring correlation functions, giving  a closed  system of equations. The steady state can now simply be calculated setting all time-derivatives equal  to zero. These algebraic equations can be solved  numerically or analytically, while filtering out the stable solution. One might try the limit $\xi\to 1 $ to approach the coupling in Section~\ref{sec:quantum bi}, only to  discover that the term $\zeta $, playing the role of a noise term, diverges.
One recovers the equations of \cite{xu_synchronization_2014}, when disregarding the relation between  $\zeta $ and  $\xi $ and setting $\zeta=\xi=1 $, which shows that the quantum coupling has no inherent coupling noise.

\subsubsection{Analysis and Comparison of the Peak Width}
Analog to Section~\ref{sec:quantum bi}  we can extract the information of the Lorentz peaks from the two-time correlation functions. They can be obtained by  using   \eqref{eq:bi-classical dynamics of s^+} and the quantum regression theorem giving a differential equation system similar to \eqref{eq:bi-quantum dynamics of expectation values}, which can be easily solved. The two-time correlation functions  consist out of  linear combinations of exponential functions  $ \Exp{- \frac 1 2  \B{\Gamma_1   \pm x_1  }\tau }$ with $x_1=\sqrt{(\gamma  N \xi   \Meansmall{\s^z})^2-\delta ^2} $ and $\Gamma_1=w   -\gamma   \B{N-1}   \Meansmall{\s^z } +    \gamma \zeta   $. The width of the Lorentz curve is nothing else but the real part $\Gamma=\Re{\Gamma_1   \pm x_1} $. To analyze the linewidth  $\Gamma $ we solve the system \eqref{eq:Aclassical bi dynamics of expectation values} up to leading order in $1/N $ giving:
 \begin{align}
\Mean{\s^z} &=\begin{cases}
                  \Min{1, \frac{w-\sqrt{w^2\xi^2-\delta^2\B{1-\xi^2}}}{N\gamma\B{1-\xi^2}} } ,&   0\leq \delta< w	\xi \\
                  \Min{1,\frac {w}{N\gamma}} ,&   \delta\geq  w\xi	
               \end{cases}
               \label{eq:Aclassical bi solution z}
.
\end{align}
Plugging \eqref{eq:Aclassical bi solution z} back into $\Gamma=\Re{\Gamma_1   \pm x_1} $  gives
\begin{align*}
\Gamma/\gamma &=   \zeta+
 \begin{cases}
                    \frac{w-\sqrt{w^2\xi^2-\delta^2\B{1-\xi^2}}}{N\gamma\B{1-\xi^2}}       ,&   0\leq \delta<w \xi	\\
                 \frac{w}{N\gamma}        ,&   \delta\geq w \xi 	
               \end{cases}
,
\end{align*}
which is valid in   the superradiant regime, i.e. upper bounded by $\Meansmall{\s^z}<1 $ using \eqref{eq:Aclassical bi solution z}.

For the plots  of the relevant functions and their analysis we refer to  Section~\ref{sec:classical bi}.

\nocite{_quantum_2016}

\end{document}